\documentclass[fleqn,usenatbib]{mnras}

\usepackage{newtxtext,newtxmath}

\usepackage[T1]{fontenc}

\usepackage{CJKutf8} 
\DeclareRobustCommand{\VAN}[3]{#2}
\let\VANthebibliography\thebibliography
\def\thebibliography{\DeclareRobustCommand{\VAN}[3]{##3}\VANthebibliography}

\usepackage{graphicx}	
\usepackage{amsmath}	

\usepackage{enumitem}
\setlist[enumerate]{leftmargin=0.6cm, labelsep=3pt, itemindent=-5pt}
\usepackage{svg}
\usepackage{float}
\usepackage{orcidlink}
\usepackage{multirow}
\usepackage[normalem]{ulem}
\usepackage{booktabs} 
\newcommand{\thispaper}{this paper}
\newcommand{\Thispaper}{This paper}
\newcommand{\NresolvedDebrisDisks}{175}
\newcommand{\ddAccessedOn}{Sept. 2024}

\newcommand{\msun}{\ensuremath{\mathrm{M}_\odot}}

\newcommand{\hmr}{\ensuremath{R_\mathrm{h}}}

\newcommand{\tcr}{\ensuremath{t_\mathrm{cr}}}
\newcommand{\trh}{\ensuremath{t_\mathrm{rh}}}
\newcommand{\tms}{\ensuremath{t_\mathrm{ms}}}

\newcommand{\tauenc}{\ensuremath{\tau_\mathrm{enc}}}
\newcommand{\nenc}{\ensuremath{\widetilde{N}_\mathrm{enc}(t)}}
\newcommand{\kmps}{\ensuremath{\mathrm{km}~\mathrm{s}^{-1}}}

\newcommand{\msunin}{\mathrm{M}_\odot}

\newcommand{\hmrin}{R_\mathrm{h}}

\newcommand{\rhocorein}{\rho_\mathrm{c}}

\newcommand{\tauencin}{\tau_\mathrm{enc}}
\newcommand{\nencin}{\widetilde{N}_\mathrm{enc}(t)}

\newcommand{\reb}{\texttt{REBOUND}}

\newcommand{\ias}{\texttt{IAS15}}

\newcommand{\nbo}{\texttt{NBODY6++GPU}}

\newcommand{\lps}{\texttt{LonelyPlanets}}

\newcommand{\fortran}{\texttt{FORTRAN}}

\newcommand{\python}{\texttt{Python}}

\newcommand{\nbody}{\textit{N}-body}
\newcommand{\lpsp}{\texttt{\textbf{LPS+}}}

\newcommand{\pdps}{test particles}

\newcommand{\Pdps}{Test particles}

\newcommand{\ensemble}{for the ensemble of all planetary systems in the star cluster}
\newcommand{\private}{\textit{private region of the planet}}
\newcommand{\reach}{\textit{reach of the planet}}
\newcommand{\territory}{\textit{territory of the planetary system}}
\newcommand{\pri}{\textit{private}}
\newcommand{\rea}{\textit{reach}}
\newcommand{\terri}{\textit{territory}}

\newcommand{\namdk}{NAMD$_{k}$}

\newcommand{\eref}[1]{equation~(\ref{#1})}
\newcommand{\Eref}[1]{Equation~(\ref{#1})}
\newcommand{\erefp}[1]{(equation~\ref{#1})}
\newcommand{\fref}[1]{Fig.~\ref{#1}}
\newcommand{\tref}[1]{Table~\ref{#1}}
\newcommand{\sref}[1]{Section~\ref{#1}}

\newcommand{\change}[1]{#1}

\title[How planets and star clusters affect comets]{Influence of planets on debris discs in star clusters - II. The impact of stellar density}

\author[Wu et al.]{
Kai Wu \begin{CJK*}{UTF8}{gbsn}(吴开)\end{CJK*}\orcidlink{0000-0003-0349-0079},$^{1,2}$
M.B.N. Kouwenhoven\orcidlink{0000-0002-1805-0570},$^{1}$\thanks{t.kouwenhoven@xjtlu.edu.cn}
Francesco Flammini Dotti\orcidlink{0000-0002-8881-3078},$^{3}$
and Rainer Spurzem\orcidlink{0000-0003-2264-7203}$^{3,4,5}$
\\
$^{1}$Department of Physics, School of Mathematics and Physics, Xi'an Jiaotong-Liverpool University, 111 Ren'ai Rd., Industrial Park District,\\ Suzhou, Jiangsu 215123, China\\
$^{2}$Department of Mathematical Sciences, University of Liverpool, Liverpool L69 3BX, UK\\
$^{3}$Astronomisches Rechen-Institut, Zentrum f\"{u}r Astronomie der Universit\"{a}t  Heidelberg, M\"onchhofstr. 12-14, D-69120 Heidelberg, Germany\\
$^{4}$National Astronomical Observatories and Key Laboratory of Computational Astrophysics, Chinese Academy of Sciences, 20A Datun Rd.,\\ Chaoyang District, 100101, Beijing, China\\
$^{5}$Kavli Institute for Astronomy and Astrophysics, Peking University, Yiheyuan Lu 5, Haidian Qu, 100871, Beijing, China\\
}

\date{Accepted ---. Received ---; in original form ---}

\pubyear{2024}

\begin{document}
\label{firstpage}
\pagerange{\pageref{firstpage}--\pageref{lastpage}}
\maketitle

\begin{abstract}
    We present numerical simulations of planetary systems in star clusters with different initial stellar densities, to investigate the impact of the density on debris disc dynamics. We use \lpsp{} to combine \nbody{} codes \nbo{} and \reb{} for simulations. We simulate debris discs with and without a Jupiter-mass planet at 50~au, in star clusters with $N=$ 1k - 64k stars. 
    The spatial range of the remaining planetary systems decreases with increasing $N$. As cluster density increases, the planet's influence range first increases and then decreases. 
    For debris particles escaping from planetary systems, the probability of their direct ejection from the star cluster decreases as their initial semi-major axis ($a_0$) or the cluster density increases. 
    The eccentricity and inclination of surviving particles increase as cluster density increases. The presence of a planet leads to lower eccentricities and inclinations of surviving particles. The radial density distribution of the remaining discs decays exponentially \change{in sparse clusters. }
    We derive a general expression of the gravitational encounter rate.
    Our results are unable to directly explain the scarcity of debris discs in star clusters. Nevertheless, given that many planetary systems have multiple planets, the mechanism of the planet-cluster combined gravitational influence on the disc remains appealing as a potential explanation.
\end{abstract} 



\raggedbottom
\section{Introduction}
\label{paper2.intro.sec}

Modern theories of planetary system formation depict a scenario in which protoplanetary discs evolve into planets \citep{zhu2021rev} and/or debris discs \citep{hughes2018ddrev}. In addition, the presence of either planets or debris discs may indicate the presence of the other. Collisions between larger objects contribute to the formation of debris discs (through a collisional cascade). The presence of debris discs indicates that 100~km size objects can form in the planetary system. Larger, kilometre-size objects may coagulate and result in the formation of planets \citep[see][and references therein]{hughes2018ddrev}.

\begin{table*}
    \caption[List of resolved debris discs in star clusters.]{List of resolved debris discs in star clusters, obtained by crossmatching the catalogue of resolved debris discs\textsuperscript{\ref{footnote:debris_disk}} with the \citet{hunt2023improving} star cluster catalogue. The first column shows the names of the star cluster to which the disc belongs. \change{The remaining columns list properties of the disc and its host star.}
    }
    \label{tab:dd_in_sc}
    \begin{tabular}{llllllllllll}
        \toprule
        \textbf{Open Cluster}       & HD     & HIP    & Spectral Type   & $L_\mathrm{bol}$ ($L_\odot$) & $T_\mathrm{eff}$ (K)  & $d$ (pc) & Age (Myr) & $R_\mathrm{disc}$ (au)  & Reference \\
        \midrule
        Melotte 25         & 28355  & 20901  & kA5hF0VmF0 & 17.6       & 7590  & 48.9  & 625       & 103 ± 28    & \citet{morales2016herschel-resolved}    \\
        HSC 1766           & 35841  &        & F3V        & 2          & 6500  & 96    & 30        & 144         & \citet{soummer2014five}                 \\
        Theia 65           & 36546  & 26062  & B8         & 20.04      & 10000 & 113.9 & $ 3 - 10 $    & $ 34 - 114 $    & \citet{currie2017subaru}        \\
        HSC 1900           & 38397  & 26990  & G0V        & 1.4        & 6000  & 55.4  & 30        & 90 ± 21     & \citet{moor2016new-debris}                   \\
        OCSN 92            & 131835 & 73145  & A2         & 9.2        & 9000  & 122   & 16        & $75 - 210$    & \citet{hung2015first}                     \\
        HSC 2931           & 146897 & 79977  & F2/F3V     & 4.2        & 6750  & 122.7 & $ 5 - 10 $    & 40          & \citet{thalmann2013imaging}                 \\
        \bottomrule
       
    \end{tabular}
\end{table*}

At the time of writing,
within over \NresolvedDebrisDisks{} resolved debris discs\footnote{\url{https://www.astro.uni-jena.de/index.php/theory/catalog-of-resolved-debris-disks.html}; accessed on \ddAccessedOn{}.\label{footnote:debris_disk}}, only 6 debris discs have been found in open clusters, and none has yet been discovered in a globular cluster (see \tref{tab:dd_in_sc} for a summary of known debris discs in star clusters). \change{The number of debris discs discovered in star clusters is significantly lower than the number of known debris discs in the Galactic field.}
Apart from observational biases, several possible causes have been proposed \citep{reiter2022YSC-PLrev}: (1) strong external photoevaporation from OB stars expels most protoplanetary discs, the predecessors of debris discs (e.g., simulations of \citealt{adams2004photoevaporation,
dai2018photoevaporation, nicholson2019rapid, cr2019b-photoevaporation}; observations in \citealt{williams2011proplydrev} and references therein; an early comprehensive review on photoevaporation \citealt{armitage2011dynamicprotorev}), and (2) debris discs may be truncated by the cumulative effect of close stellar encounters. 

In star clusters, frequent stellar flybys affect the stability of planetary systems. 
This has been verified using computer simulations utilising various computational methods, such as 
(i) planetary systems where the data on stellar flybys is derived from theoretical distributions \citep[e.g.,][]{malmberg2007close, malmberg2011the-effects, hao2013the-dynamical, hamers2017hot-jupiters, li2019fly-by}, 
(ii) planetary systems containing one or two planets that are simulated in star cluster codes \citep[e.g.,][]{spurzem2009planets, liu2013configurations, zheng2015the-dynamical, shara2016dynamical, van-elteren2019survivability, flamminidotti2023proplyd}, 
and (iii) multi-planet systems within star clusters that are modelled using the \lps{} Scheme and its variations \citep[e.g.,][]{cai2017stability, cai2019on-the-survivability, flamminidotti2019planetary, stock2020resonant, wu2023influence, benkendorff2024hot-jupiter}.
These studies show that close encounters shape the architectures of planetary systems. Some planets are ejected from their host systems, while others remain bound, often changing their orbital energy and/or angular momentum.

A number of earlier studies on the dynamical evolution of circumstellar discs with single close stellar encounters have been carried out \citep[see, e.g.,][and references therein]{olczak2006encounter-triggered,umbreit2011disks,cuello2023single-encounter-rev}. Recent studies have started to explore the dynamics of planetesimal discs directly in the context of star cluster simulations \citep[e.g.,][]{hands2019the-fate, veras2020linking, cai2019on-the-survivability, wu2023influence}. 

Only a small fraction of the known debris discs are located in star clusters. None of the six debris-hosting stars in star clusters (\tref{tab:dd_in_sc}) is known to host planets yet. This is likely the result of observational limitations that prevent the detection of such systems, but may also partially be attributed to the gravitational influence of the planets and the stellar flybys.

Planets play an important role in shaping debris discs. Previous investigations of isolated planetary systems show that the structure of debris discs is often highly sensitive to the masses, semimajor axes, and eccentricities of planets \citep[e.g.,][]{nesvold2015a-smack, nesvold2016circumstellar, tabeshian2016detection, tabeshian2017detection, zheng2017clearing}. Such effects can be more complex when planet migration is considered \citep{marzari2013circumstellar,cattolico2020dynamical,wu2023influence}. In dense stellar environments, close stellar encounters excite the orbits of planets, which may lead to ejections or physical collisions, and a change in the structure of the debris disc. 

The joint effect of planets and star cluster environment on the evolution of circumstellar debris discs is complex. \citet{wu2023influence} studied such systems in intermediate-mass ($N=8000$~stars) star clusters and found that the introduction of a Jupiter-mass planet at 50~au destabilizes the debris discs in the star cluster. The planet removes all particles with $40~\mathrm{au}<a<60~\mathrm{au}$ and affects particles with $a<400$~au. The wide range in semi-major axes of the affected debris particles can be attributed to perturbed planetary orbits. As a consequence of stellar encounters, debris discs in such environments are typically truncated \change{at 700~au}, and debris discs extending beyond this range are rare. Denser star cluster environments may truncate debris discs further.

It is possible that planets, particularly giant planets, may be less likely to form in dense stellar environments \citep{ndugu2018planet, ndugu2022planet}. Nevertheless, given the large number of stars in each star cluster, and the likelihood that most field stars originate from dense stellar environments, it is important to study the formation and evolution of planetary systems in star clusters. In this study, we use gravitational \nbody{} simulations to investigate how debris discs dynamically evolve in star clusters with different initial stellar densities. We focus on the impact of the presence (or absence) of a planet in the system, to explore the evolution of the planetary systems hosting planets and debris discs in star clusters. Our methodology is based on an improvement of the \lps{} scheme \citep{cai2017stability, cai2019on-the-survivability, flamminidotti2019planetary, stock2020resonant}.

\Thispaper{} is organized as follows. \sref{paper2.method.sec} describes the numerical methods and initial conditions. \sref{paper2.results.sec} presents the results. Finally, we summarize and discuss our conclusions in \sref{paper2.conclusions.sec}.

\section{Methodology and initial conditions}
\label{paper2.method.sec}

\subsection{Numerical approach}
\label{paper2.code.sec}

\begin{figure} \centering
    \includegraphics[width=0.47\textwidth]{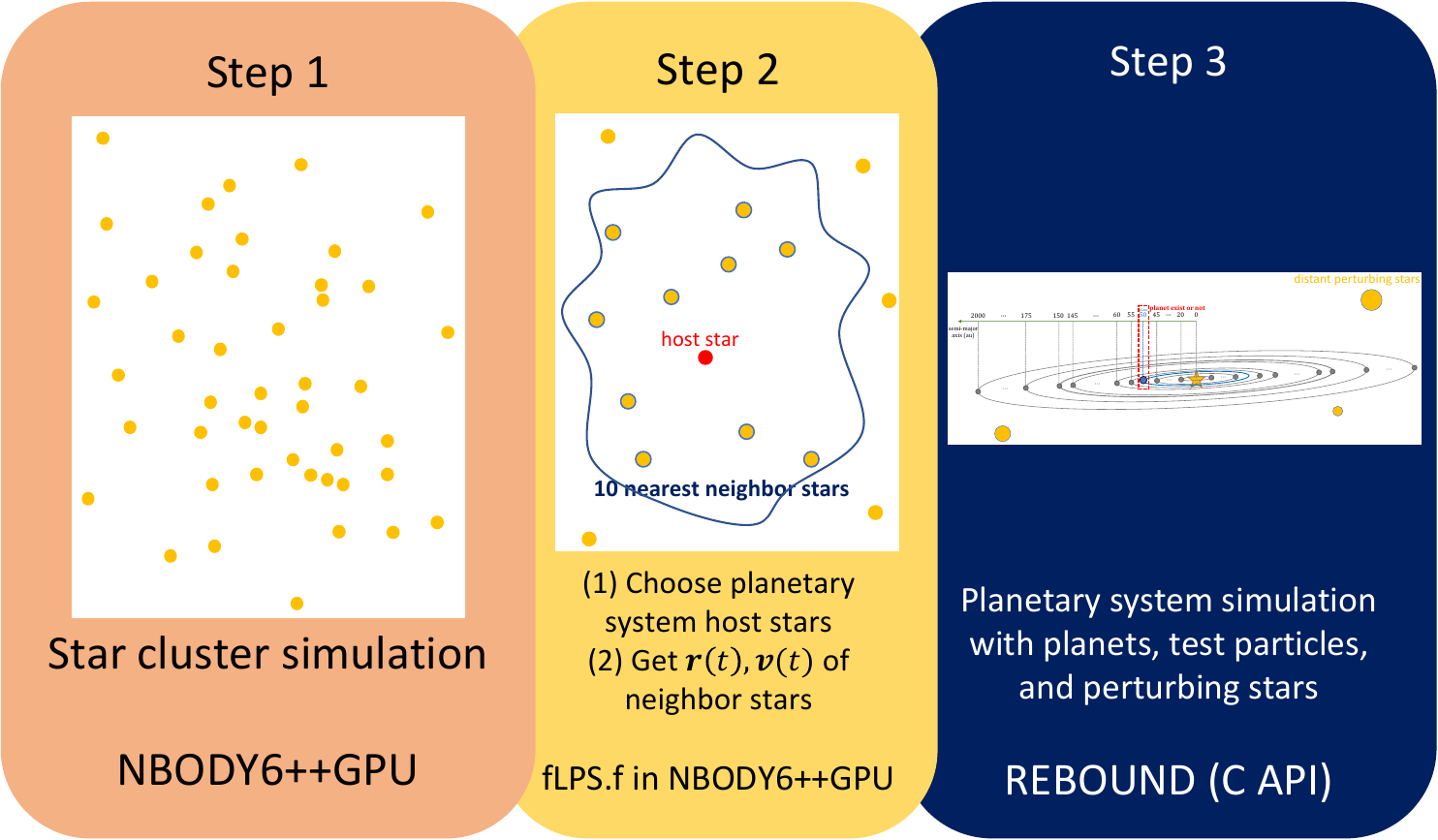} 
    \caption[Schematic diagram of the computational approach \lpsp{}.]{Schematic diagram of the computational approach \lpsp{} \citep{wu2023influence}. A high-resolution version of Step~3's schematic diagram can be obtained by zooming in on the digital version of \thispaper{}, and can also be found in figure.~2 of \citet{wu2023influence}.}
    \label{fig.method}
\end{figure}

We adopt the computational methodology of \citet{wu2023influence}, summarized in \fref{fig.method}. We refer to this code as \lpsp{}. We first evolve star clusters using \nbo{}\footnote{\url{https://github.com/nbody6ppgpu/Nbody6PPGPU-beijing}} \citep{NBODY6++GPU,NBODY6++GPUNewSSE,spurzem2023rev}. Stellar evolution of level C is enabled, following the prescriptions described in \citet{NBODY6++GPUNewSSE}. We use a modified version of \nbo{}, in which we insert the \fortran{} routine \texttt{fLPS.f} to (i) initially, select 100 solar mass stars as the host stars of planetary systems, \change{(ii) at each output timestep, we identify for each host star the ten nearest stars (i.e., perturbers),} and (iii) store the kinematic data of these perturbers at high time resolution (which can reach an average output interval of 50 yr, i.e., up to 20\,000 snapshots per Myr) for the subsequent simulations of the planetary systems. 

We integrate the individual planetary systems using \reb{}\footnote{\url{https://github.com/hannorein/rebound}} \citep{rein2012rebound} with the \ias{} integrator \citep{rein2015ias15}. Each of these simulations includes the host star, a planet, 200 test particles, and ten nearest neighbour stars. We refer to \citet{wu2023influence} for a detailed description of the adopted computational approaches. 

\change{Following the approach adopted by \cite{cai2019on-the-survivability,flamminidotti2019planetary,stock2020resonant} and \cite{benkendorff2024hot-jupiter}, a} particle is marked as having escaped from the planetary system when its eccentricity exceeds $e=0.99$. \change{Physical collisions between the bodies in the planetary systems do not occur. None of the particles experience sufficiently close encounters.}

\subsection{Initial conditions}

\subsubsection{Initial conditions of star clusters}

\begin{table}
    \caption{Initial conditions of the star cluster models.}
    \label{tab.scini_comm}
    \centering
    \begin{tabular}{ll}
        \toprule
        Quantity & Value \\
        \midrule
        Initial number of stars    & $1\,000-64\,000$ (see \tref{tab.scini_diff}) \\
        Initial density profile    & \citet{plummer1911profile}   \\
        Initial mass function      & \citet{kroupa2002the-initial}; $0.08 - 150$~\msun{}  \\
        Initial half-mass radius   & $\hmrin{}=0.78$~pc            \\
        Initial virial ratio       & $Q=0.5$     \\
        Tidal field                & Solar orbit in the Milky Way \\
        Primordial binaries        & None        \\
        Metallicity                & $Z=0.001$  \\
        Stellar evolution          & Enabled      \\
        Integration time           & 100~Myr \\
        \bottomrule
    \end{tabular}
\end{table}

\begin{table}
    \caption[Initial properties of the star cluster models.]{Initial properties of the star cluster models. $N$: initial number of stars. $M$: initial total cluster mass. \tcr{}: initial crossing time. \trh{}: initial half-mass relaxation time. \tms{}: initial mass segregation time scale for solar-mass stars. $M_\textrm{host}$: mass range of host stars of planetary system simulations. The data for the 8k model data are identical to those in \citet{wu2023influence}.}
    \label{tab.scini_diff}
    \centering
    \begin{tabular}{ccccccc}
    \toprule
    ID  & $N$      & $M$          & \tcr{}      & \trh{}      & \tms{}  & $M_\textrm{host}$        \\ 
        &          & (\msun{})    &  (Myr)      & (Myr)       & (Myr)   & (\msun{})                \\ \midrule
    1k  & ~1\,000  & 628          & 0.5622      & 9.157       & 5.747   & [0.622, 1.354]           \\
    2k  & ~2\,000  & 1119         & 0.3929      & 11.47       & 6.416   & [0.812, 1.192]           \\
    4k  & ~4\,000  & 2230         & 0.3020      & 15.98       & 8.910   & [0.892, 1.108]           \\
    \textit{8k}  & \textit{8\,000}  & \textit{4662}         & \textit{0.2089}      & \textit{20.21}    & \textit{11.78}   & \textit{[0.981, 1.015]}           \\
   
    16k & ~16\,000 & 9507         & 0.1532      & 27.29       & 15.48   & [0.992, 1.009]           \\
    32k & ~32\,000 & 18919        & 0.1058      & 34.93       & 20.22   & [0.995, 1.004]           \\
    64k & ~64\,000 & 38079        & 0.0761      & 46.82       & 27.32   & [0.998, 1.002]           \\ 
    \bottomrule
    \end{tabular}
\end{table}

The initial conditions of the star cluster models are summarized in \tref{tab.scini_comm}. Inspired by related studies \citep{flamminidotti2019planetary,stock2020resonant,veras2020linking,wu2023influence}, we simulate star clusters with different numbers of stars, ranging from $N=1\,000$ to $N=64\,000$. We model seven star clusters in total, including the 8k model of \citet{wu2023influence}. The initial properties of these models are summarized in \tref{tab.scini_diff}. 

We adopt \citet{plummer1911profile} model for all clusters. The central density of a Plummer sphere is \citep{heggiehut2023-million-body-book}
\begin{equation}
    \rhocorein{} = \frac{3 (2^{2/3}-1)^{-3/2}  }{4\pi}    \frac{M}{\hmrin{}^3} 
    \approx 0.53 \frac{M}{\hmrin{}^3} 
    = 0.53\frac{\bar{m}_{*} N}{\hmrin{}^3} \quad ,
    \label{eqn.core_density}
\end{equation}
where $M$ is the total cluster mass, \hmr{} is the half-mass radius, and $\bar{m}_{*}$ is the average stellar mass.
We adopt an initial mass function (IMF) of \citet{kroupa2002the-initial} with a minimum mass of $0.08$~\msun{} and a maximum mass of $150$~\msun{}, \change{which corresponds to an average mass of $\bar{m}_{*} \approx 0.57~\msunin{}$ for all cluster models}. We adopt identical initial half-mass radius ($\hmrin{}=0.78$~pc), so that the central density in each star cluster scales as $\rhocorein \propto N$.
The 8k - 64k models have very similar central density as those in \citet{stock2020resonant}. We thus refer to fig.~1 of \citet{stock2020resonant} for a comparison of the density of the models with observed clusters. Our 1k, 2k, and 4k models have central densities of 57.82, 65.32, 77.51 $\msunin{}~\mathrm{pc}^{-3}$ at 100~Myr, which is comparable to that of Pleiades \citep{raboud1998investigation,marks2012inverse,fujii2019survival}.

The star clusters are initially in virial equilibrium. We adopt an external tidal field corresponding to that of the solar orbit in the Milky Way. For simplicity, we do not include primordial binaries{, triples, or higher-order multiples. The limitation of the zero-binary assumption is discussed in \sref{paper2.conclusions.sec}. 

All models start with single-population zero-age main sequence stars. An integration time 100~Myr is sufficient for the purpose of our study in the 8k model, as discussed in \citet{wu2023influence}. In \sref{sec.result_star_cluster}, we discuss whether the 100~Myr simulation time is sufficient for debris discs to evolve into a state of pseudo-equilibrium in the other models. 

In each star cluster, we select a hundred stars with masses of $\sim 1$~\msun{} as host stars \citep[see also][]{cai2017stability, cai2019on-the-survivability, flamminidotti2019planetary, veras2020linking, wu2023influence}. Due to the sampling of the stellar population from the IMF, host star masses deviate somewhat from 1~\msun{} (see \tref{tab.scini_diff}). All models with $N \ge 8\,000$ have $M_\textrm{host}$ within 2\% difference from 1~\msun{}, but for small star clusters, the mass range of host stars is broader. 

\subsubsection{Initial conditions of planetary systems}

\begin{table}
    \caption{Initial conditions of the planetary systems.}
    \label{tab.plini}
    \begin{tabular}{lll}
        \toprule
        Property               & Planet      & \Pdps{}     \\
        \midrule                                             
        Number                 & 0 or 1       & 200        \\
        Mass                   & $m_p= 1 M_\mathrm{J} = 317.83 M_\oplus$ & $m_c=0$       \\
        Semi-major axis        & $a_p=50$~au     & $a_c=20-2000$~au    \\
        Eccentricity           & $e_p= 0.0484$ ($e_\mathrm{J}$)   & $e_c=0.01$             \\
        Inclination            & $i_p= 0^\circ$ (reference plane)  & $i_c = 0.01~\mathrm{rad}=0.573^\circ$ \\
       
        \bottomrule
    \end{tabular}
\end{table}

To investigate the influence of the star cluster density, we adopt the same initial conditions of planetary systems as \citet{wu2023influence} for all models, summarized in \tref{tab.plini}. \change{Between 20~au and 150~au, we place \pdps{} at intervals of $\Delta a_0=5$~au (except at $a_0 = 50$~au, where a planet may be located). Between 150~au and 2000~au, we place \pdps{} at intervals of $\Delta a_0=25$~au.} This results in 100 different $a_0$. On each $a_0$ we place two particles with random periastrons and ascending nodes. Hereafter, models with planets are referred to as \textbf{P1}, and those without planets as \textbf{P0}. Both P0 and P1 simulations contain a hundred planetary systems in which the host stars have masses of $M\approx 1$\,\msun. We include ten perturbers when modelling the planetary systems. Planets and test particles are removed when the eccentricity exceeds $e=0.99$. We refer to \citet{wu2023influence} for a detailed motivation of initial conditions of planetary systems. 

\change{Restricted by the current limitations of the simulation code \citep{wu2023influence}, we are unable to follow the trajectories of the particles after they escape from their planetary systems. This is because planetary system simulations are conducted after completion of the star cluster simulation. In future studies, we will focus on implementing the ability to track capture and recapture events of planetary particles by other stars in \lpsp{}. }

\section{Results}
\label{paper2.results.sec}

\subsection{Theoretical background}
\label{sec.encounter_timescale_theory}

\begin{figure} \centering
    \includegraphics[width=0.47\textwidth]{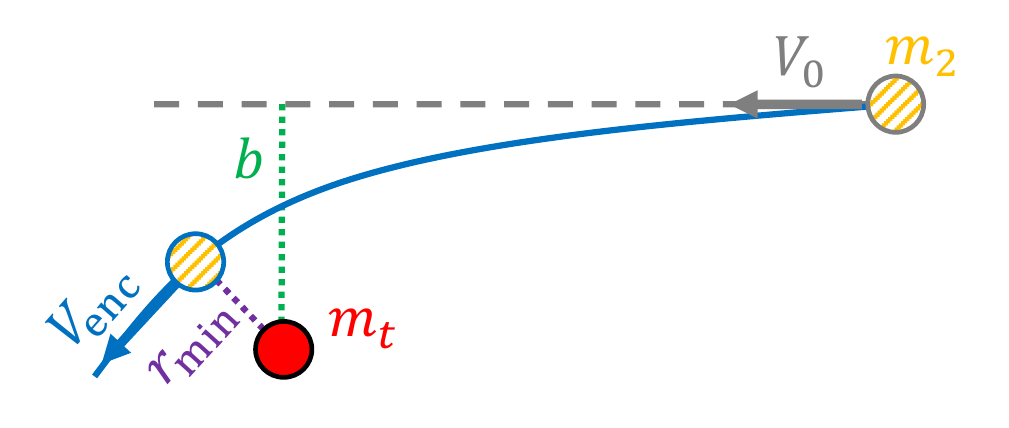} 
    \caption{Schematic diagram of a stellar close encounter.}
    \label{fig.encounter_schem}
\end{figure}

Before we analyse the simulation data, we first derive general expressions for the encounter rates and encounter timescales of planetary systems in different star cluster environments, without assuming the domination of gravitational focusing. 

The encounter rate, $f_\mathrm{enc}$, at which a star experiences encounters within a distance $r_{\min}$ (see \fref{fig.encounter_schem}) with other stars is
\begin{equation}
    f_\mathrm{enc} = \frac{1}{\tauencin{}} = n V_\mathrm{0} \Sigma \quad ,
\end{equation}
where \tauenc{} is the encounter timescale, $n$ is the number density of stars, $V_\mathrm{0}$ is the relative velocity of the two stars long before their encounters, and $\Sigma$ is the cross-section including the gravitational focusing.

A typical value for $n$ is the average number density within the half-mass radius, $\hmrin$, of the star cluster, 
\begin{equation}
    n \approx \frac{ \frac{1}{2} N }{ \frac{4}{3} \pi \hmrin{}^3 } 
      = \frac{3}{8 \pi \hmrin{}^3} N 
      = \frac{3}{8 \pi \hmrin{}^3}\frac{M_\mathrm{cluster}}{\bar{m}_{*}} \quad ,
\end{equation}
where $\bar{m}_{*}$ is the average stellar mass in the cluster. 

We have $V_\mathrm{0} = \sqrt{6}\sigma_\mathrm{1d}$ as the 3D dispersion of relative velocity for particles following Maxwellian distribution \citep[e.g.,][exercise 4.18]{galacticDynamics2ndEdition}, where $\sigma_\mathrm{1d}$ is the one-dimensional velocity dispersion of stars. The $\sigma_\mathrm{1d}$ within the half-mass radius in the Plummer model \citep{plummer1911profile} can be expressed \citep{heggiehut2023-million-body-book} by
\begin{equation}
    \begin{aligned}
        \sigma_\mathrm{1d} 
        &= \sqrt{-\frac{1}{6} \phi(\hmrin{})} = \sqrt{\frac{k \left(1+k^2\right)^{-\frac{1}{2}}}{6} \frac{G M_\mathrm{cluster}}{\hmrin{}}} \\
        &= \left( C~\frac{ G M_\mathrm{cluster} }{ \hmrin{} } \right)^{1/2} \quad ,
    \end{aligned}
\end{equation}
where $k=\sqrt{1/(2^{2/3}-1)} \approx 1.305$ is the ratio of half-mass radius to Plummer radius, and constant $C = k\,(1+k^2)^{-1/2}/6 \approx 0.1323$. \change{Initially, $\sigma_\mathrm{1d}$ is 1.84~\kmps{} in the 8k model. Initial values for the other models range from 0.68~\kmps{} (for the 1k model) to 5.27~\kmps{} (for the 64k model).}

The cross-section is $\Sigma = \pi b^2$, where $b$ is the impact parameter. We derive the expression for $b$ below.
During an encounter between two stars, the energy is conserved
\begin{equation}
    \begin{split}
         \frac{1}{2} \mu V_\mathrm{0}^2 = E_0 = \frac{1}{2} \mu V_\mathrm{enc}^2 - \frac{G m_\mathrm{t} m_2}{r_\mathrm{min}} \quad ,
    \end{split}
    \label{eqn.E}
\end{equation}
where $\mu=m_\mathrm{t}m_2/(m_\mathrm{t}+m_2)$ is the reduced two-body mass, $V_\mathrm{enc}$ is the relative velocity of the two stars at their closest approach, $m_\mathrm{t}$ is the mass of the target star which we evaluate encounter, $m_\mathrm{2}$ is the mass of the other star involved in the encounter, and $r_\mathrm{min}$ is the minimum distance (the periastron distance) of the two stars in the encounter. Angular momentum conservation gives the relation
\begin{equation}
    \begin{split}
        \mu b V_\mathrm{0} = L_0 = \mu r_\mathrm{min} V_\mathrm{enc} \quad .
    \end{split}
    \label{eqn.L}
\end{equation}
Combining \eref{eqn.E} and (\ref{eqn.L}) and eliminating $V_\mathrm{enc}$, we obtain
\begin{equation}
    \begin{split}
        b^2 = r_\mathrm{min}^2 + \frac{ 2G (m_\mathrm{t} + m_\mathrm{2}) r_\mathrm{min} }{ V_\mathrm{0}^2 } = r_\mathrm{min}^2 \left[ 1 + \frac{ 2G (m_\mathrm{t} + m_\mathrm{2}) }{ V_\mathrm{0}^2 r_\mathrm{min}} \right]  \quad .
    \end{split}
    \label{eqn.cross-section}
\end{equation}

The second term between the brackets of \eref{eqn.cross-section} denotes the gravitational focusing's contribution to the cross-section, $\eta_\mathrm{focus}$, which is
\begin{equation}
    \eta_\mathrm{focus} \equiv \frac{ 2G (m_\mathrm{t} + m_\mathrm{2}) }{ V_\mathrm{0}^2 r_\mathrm{min}} = \frac{(m_\mathrm{t} + m_\mathrm{2}) ~ \hmrin{}}{ C~M_\mathrm{cluster} ~ r_{\min}} \quad .
    \label{eqn.eta-focus}
\end{equation}

In our models, we consider encounters experienced by stars of mass $m_\mathrm{t} = 1~\msunin{}$ in clusters with $\hmrin{} = 0.78$~pc initially. If we evaluate encounters within $r_{\min} = 1\,000$~au, and assume equal-mass encounter $m_\mathrm{t} = m_\mathrm{2}$, then $\eta_\mathrm{focus}$ ranges from 3.87 (1k model) to 0.06 (64k model) at the start of simulations. This implies that it is not appropriate to assume that gravitational focusing dominates (i.e., $\eta_\mathrm{focus} \gg 1$) when simulations start. The densest 64k cluster has $\eta_\mathrm{focus} \ll 1$, which means gravitational focusing is negligible. Thus, we cannot assume $\eta_\mathrm{focus} \gg 1$ for our models. 

The assumption that encounters are dominated by gravitational focusing is acceptable in several situations in our star cluster models. Firstly, when only very close encounters are evaluated, i.e., when $r_{\min}$ is small, gravitational focusing dominates stellar encounters. For example, the choice of $r_{\min} = 100$~au results in 10 times higher value of $\eta_\mathrm{focus}$ than that of $r_{\min} = 1\,000$~au. When $r_{\min} = 100$~au,  the 1k model at $t=0$~Myr has $\eta_\mathrm{focus,100au}=38.7$. The choice of $r_{\min}$ depends on the purpose of each study. Secondly, gravitational focusing may dominate the encounters when a sparse cluster has evolved for several relaxation times. For instance, our sparsest 1k cluster has $\hmrin{}=3.5$~pc and $M_\mathrm{cluster}=364~\msunin{}$ at 100~Myr, and thus $\eta_\mathrm{focus}=30.0$. In the latter case, it may be acceptable to adopt $\eta_\mathrm{focus} \gg 1$.

Finally, the general expression of the encounter rate of planetary systems in star clusters can be written as
\begin{equation}
    \begin{split}
        f_\mathrm{enc}= \frac{3\sqrt{6CG}}{8} 
            M_\mathrm{cluster}^{3/2}~ 
            & \bar{m}_{*}^{-1}~ \hmrin{}^{-7/2}~ 
            r_{\min}^2 \left[1+ \frac{(m_\mathrm{t} + m_\mathrm{2}) ~ \hmrin{}}{ C~M_\mathrm{cluster} ~ r_{\min}}\right] \quad .
    \end{split}
    \label{eqn.deduct_encounter_rate}
\end{equation}
We apply convenient units to \eref{eqn.deduct_encounter_rate}, which results in
\begin{equation}
    \begin{split}
    f_\mathrm{enc} = 1.67 \times 10^{-2}
        &~\mathrm{Myr}^{-1}
        \left( \frac{M_\mathrm{cluster}}{1000~\msunin{}} \right)^{3/2}
        \left( \frac{\bar{m}_{*}}{1~\msunin{}} \right)^{-1}
        \left( \frac{\hmrin{}}{1~\mathrm{pc}} \right)^{-7/2} \\
        & \times \left( \frac{r_{\min }}{10^3~\mathrm{au}} \right)^2
        \left[1+ \frac{(m_\mathrm{t} + m_\mathrm{2}) ~ \hmrin{}}{ 0.1323~M_\mathrm{cluster} ~ r_{\min}} \right] \quad ,
    \end{split}
    \label{eqn.encounter_rate}
\end{equation}
and the close encounter timescale is $\tauencin{} = 1/f_\mathrm{enc}$~. 
The proportionality between \tauenc{} and $M_\mathrm{cluster}$ depends on the degree of gravitational focusing. In the case of negligible gravitational focusing (i.e., geometric encounter, $\eta_\mathrm{focus} \ll 1$), the relationship is $\tauencin{} \propto M_\mathrm{cluster}^{-3/2} \propto N^{-3/2}$. When gravitational focusing dominates the encounter ($\eta_\mathrm{focus} \gg 1$), the relationship becomes $\tauencin{} \propto M_\mathrm{cluster}^{-1/2} \propto N^{-1/2}$.

Note that \eref{eqn.encounter_rate} provides a general estimate of the close encounter rate. It comes with several assumptions: (1) the cluster is assumed to be a Plummer sphere, (2) the number density is represented by that within the half-mass radius, and (3) the velocity-at-infinity of the encountering stars is represented by average velocity dispersion within the half-mass radius. \Eref{eqn.encounter_rate} provides a good expression to estimate encounter rates in star clusters. The limiting case in which gravitational focusing dominates ($\eta_\mathrm{focus} \gg 1$), such as the expressions used in \citet{malmberg2007close,maraboli2023encounters}, have already been well-tested. 

\subsection{Star cluster evolution}
\label{sec.result_star_cluster}

\begin{figure} \centering
    \includegraphics[width=0.44\textwidth]{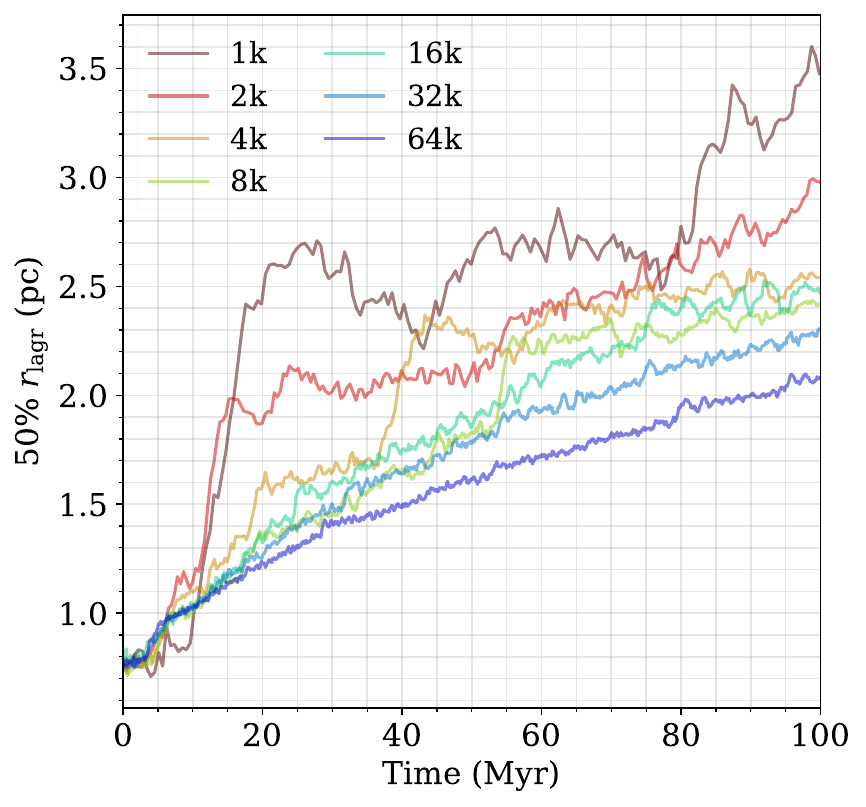}
    \includegraphics[width=0.44\textwidth]{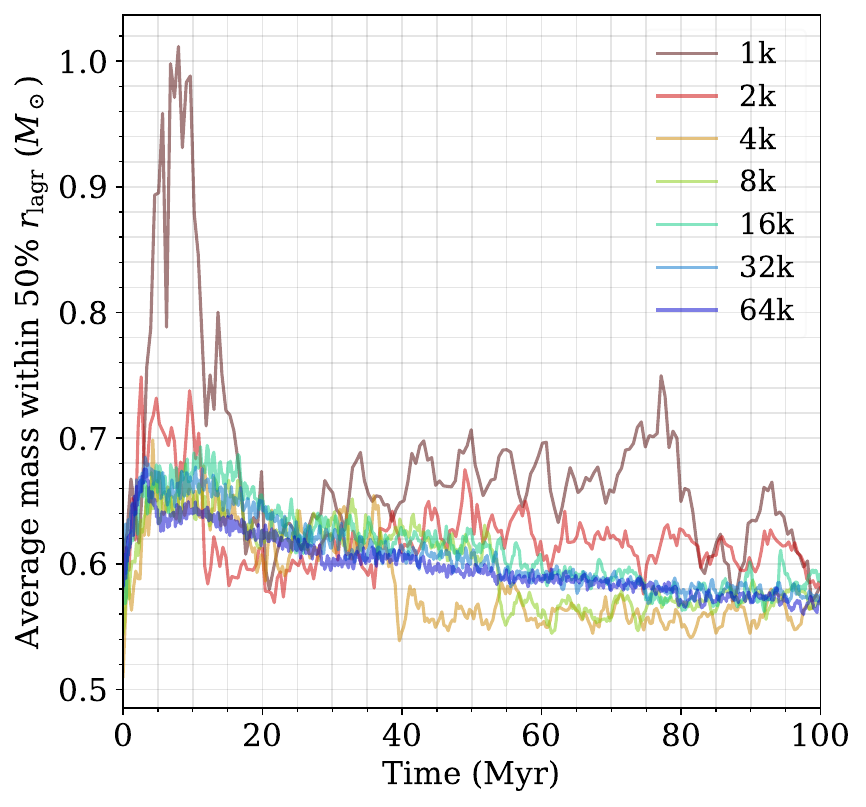}
    \includegraphics[width=0.44\textwidth]{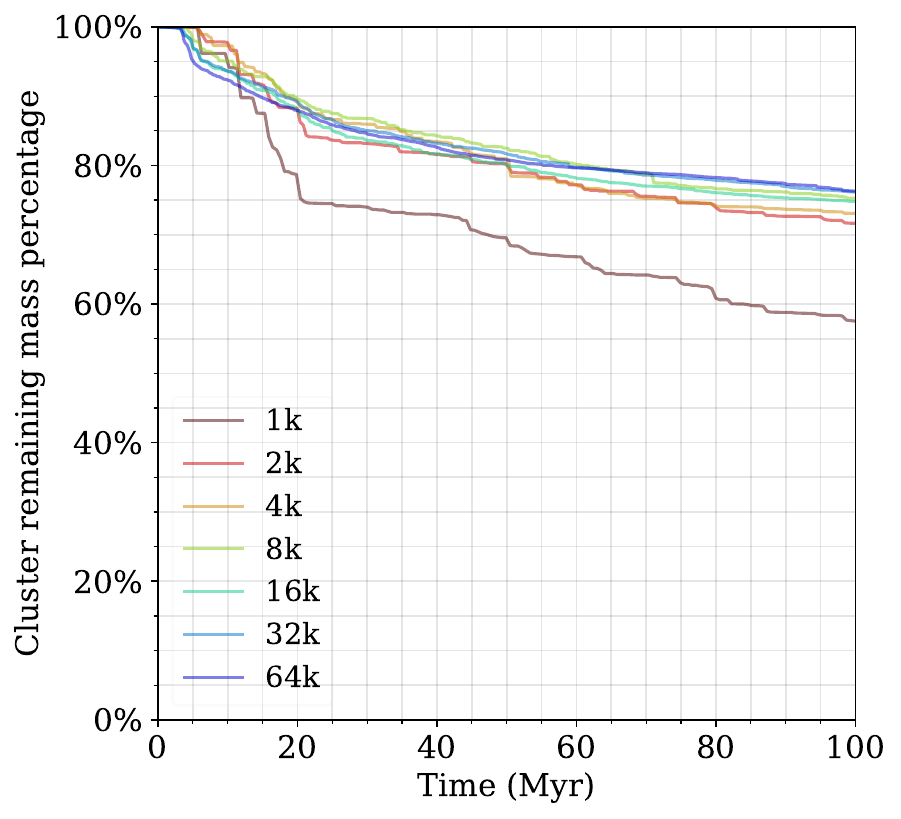} 
    \caption[Evolution of the 50\% Lagrangian radii, of the average stellar mass within the 50\% Lagrangian radii, and of the remaining cluster mass relative to the initial mass for different star clusters models.]{Evolution of the 50\% Lagrangian radii (top), of the average stellar mass within the 50\% Lagrangian radii (middle), and of the remaining cluster mass relative to the initial mass (bottom) for different star clusters models.}
    \label{fig.star_cluster_evo}
\end{figure}

\begin{figure} \centering
    \includegraphics[width=0.44\textwidth]{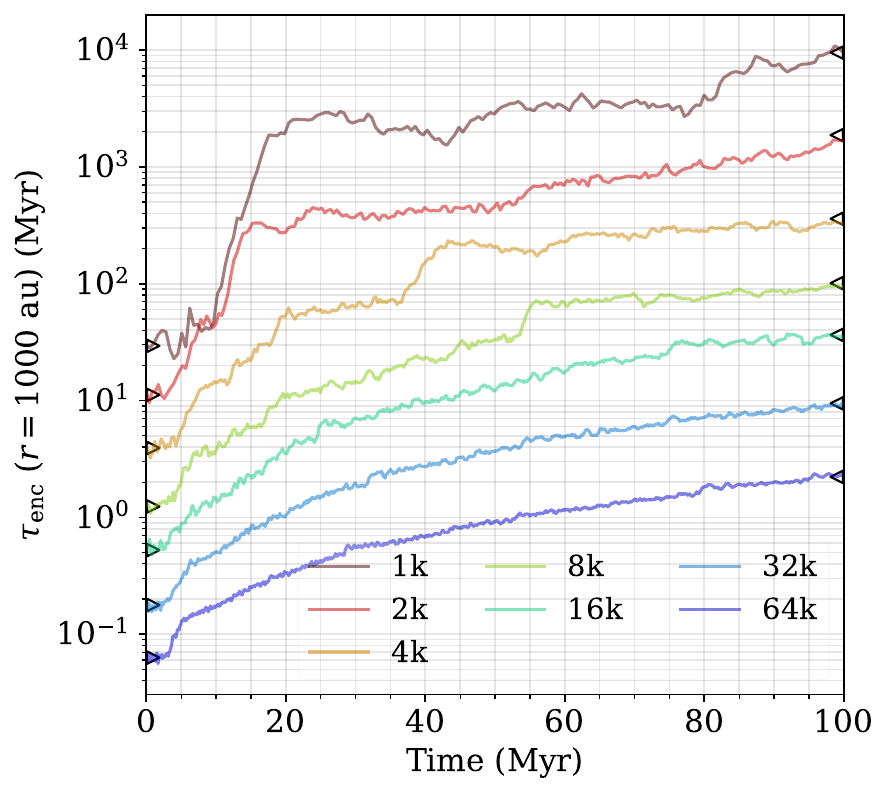} 
    \caption[Evolution of the encounter timescale, \tauenc{}.]{Evolution of the encounter timescale, \tauenc{}, estimated using \eref{eqn.encounter_rate} with $m_2 = \bar{m}_{*}$ and $r_\mathrm{min} = 1\,000$~au. We fit \tauenc{} as a function of $N$ with a power-law function. The optimal fits yield that $\tauencin$ scales as $N^{-1.49}$ at $t=0$, and as $N^{-1.58}$ at $t=100$~Myr. The right- and left- pointing triangles correspond to the calibrated values obtained from the power-law fitting at $t=0$ and 100~Myr, respectively.
    }
    \label{fig.tau_enc_evolution}
\end{figure}

\begin{figure} \centering
    \begin{tabular}{c}
        \includegraphics[width=0.44\textwidth]{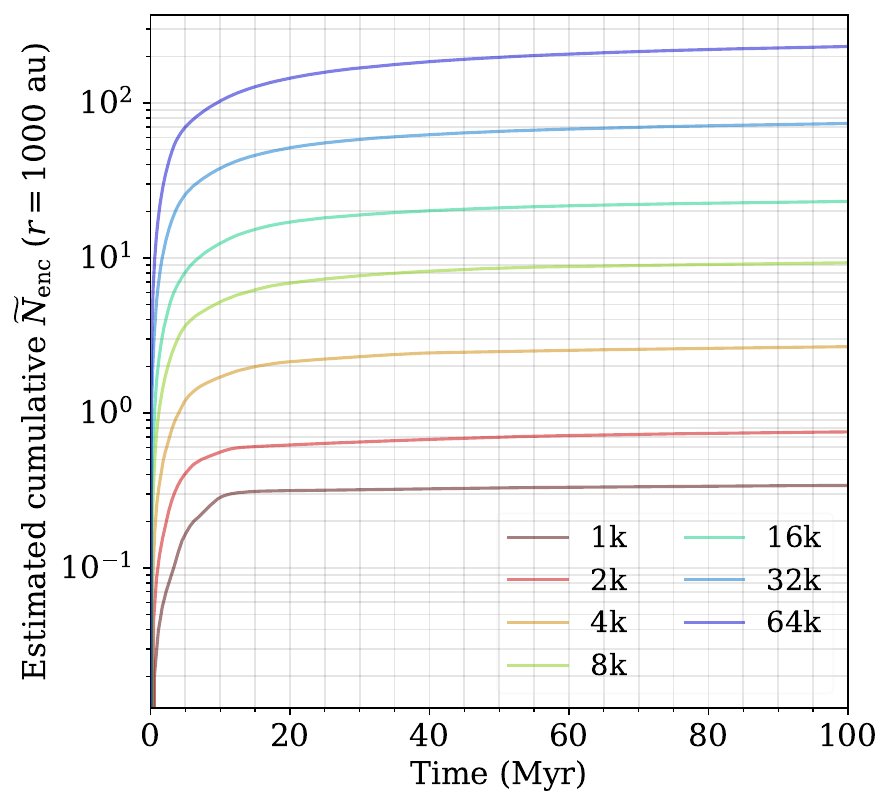} \\
        \includegraphics[width=0.44\textwidth]{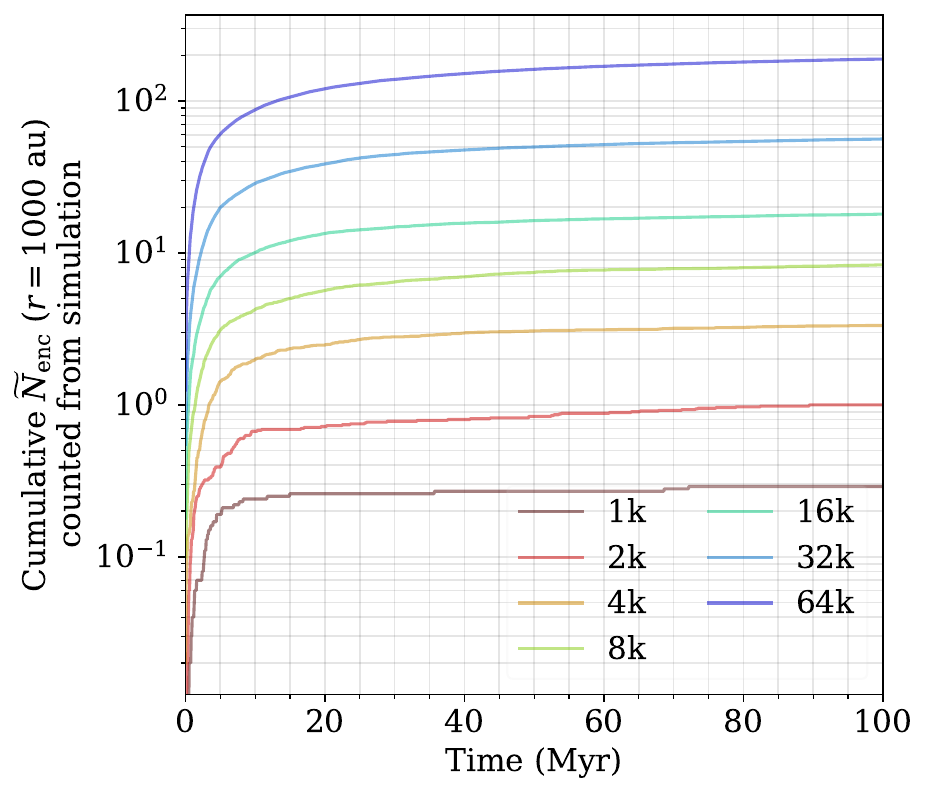}
    \end{tabular}
    \caption[Cumulative number of encounters per host star, $\widetilde{N}_\mathrm{enc}$]{Cumulative number of encounters per host star, $\widetilde{N}_\mathrm{enc}$. \change{Upper panel: values estimated using \eref{eqn.n_enc}. Lower panel: values measured in simulations.}}
    \label{fig.nenc}
\end{figure}

The close encounter rate experienced by the planet-hosting star changes over time as the star clusters evolve. \fref{fig.star_cluster_evo} shows substantial changes in the total cluster mass, $M_\mathrm{cluster}$, and the half-mass radius, \hmr{}, of each cluster. The average stellar mass within the 50\% Lagrangian radius, $\bar{m}_{*}$, shows an intermittent increase of up to $\sim 1$\,\msun{} in the sparsest cluster and of $\sim 0.68$\,\msun{} in the densest cluster.
\change{Less than half of the mass loss in each star cluster is attributable to stellar ejections: $42\%-15\%$ for models 1k$-$64k, respectively. The remainder of the mass loss is a consequence of stellar evolution.}

Close encounter can be destructive to a planetary system. We evaluate \tauenc{} within $r_\mathrm{min} = 1\,000$~au.
In the estimates below, we consider the average stellar mass ($m_2 = \bar{m}_{*}$ in equation~\ref{eqn.encounter_rate}). A more accurate estimate for encounter rate can be obtained using statistics of perturbing star mass, if desired. 

\fref{fig.tau_enc_evolution} shows the evolution of encounter timescales, \tauenc{}, where time-dependent values of $\bar{m}_{*}$, \hmr{} and $M_\mathrm{cluster}$ are adopted. The encounter timescales increase with cluster age. Different relationships between \tauenc{} and $N$ indicate different levels of gravitational focusing (\sref{sec.encounter_timescale_theory}). In \fref{fig.tau_enc_evolution}, the relationship between \tauenc{} and the initial stellar number in the cluster $N$ is roughly $\tauencin{} \propto N^{-3/2}$ both the start and the end of the simulations. This suggests that gravitational focusing plays a negligible role in all our star cluster models, including the sparest 1k cluster.

Since $f_\mathrm{enc} = \tauencin{}^{-1}$ represents the number of encounters per unit of time, we can estimate the cumulative number of encounters that each planetary system experiences in a time interval $t$,
\begin{equation}
    \nencin{} = \int_{0}^{t} f_\mathrm{enc}(T) \,dT
    \quad .
    \label{eqn.n_enc}
\end{equation}

\fref{fig.nenc} shows \nenc{} for the different models. \change{We compare the predicted number of encounters using \eref{eqn.n_enc}, with the actual number of encounters in each simulation. The latter is obtained by counting all events in which the distance between a perturbing star and its host star is $r\leq 1000$~au. The number of encounters obtained from estimation and the simulation data are in good agreement, differing less than 17\% at all times.
In all simulations, the encounter rate per host star has dropped significantly by $t=100$~Myr.}

We estimate the time interval between two subsequent encounters at $t\approx 100$~Myr by extrapolating curves in \fref{fig.nenc}. Cluster models 1k - 8k are relatively sparse. It typically takes at least another 100~Myr for planetary systems in these systems to experience an additional encounter. For the 32k and 64k models, the next encounter is expected to occur roughly after 20~Myr and 10~Myr, respectively. Longer simulations will be required to study the long-term evolution of planetary systems in the clusters (16k, 32k, and 64k). Here, we choose to limit our simulations to 100~Myr, which allows us to compare our results with those of previous studies \citep[e.g.,][]{cai2019on-the-survivability,veras2020linking,stock2020resonant}. The relaxation time of a star cluster increases as the cluster evolves. Consequently, all models effectively only experience one or two relaxation times. Note, however, that star clusters may also experience core collapses at later times, which may result in periods of high close-encounter rates at times beyond 100~Myr. 

\citet{zheng2015the-dynamical} studied star clusters with $N=1000$, in which all stars are initially assigned a planetary companion. They consider the full spectrum of host star masses. After dynamically evolving for 50~Myr, 3.5\% of the host stars, along with their planetary systems (\textit{EPSs} in their paper), have escaped from the star cluster. The fraction of planetary systems escaping from the star cluster is potentially a function of host star mass, average stellar mass, and stellar density \citep[see][for a recent review of stellar escape mechanisms]{weatherford2023escape-star}. In our simulations, a small fraction of their host stars escape from the star cluster along with their planetary systems. The fractions are 18\%, 6\%, 5\%, 3\%, 1\%, 1\%, and 2\% in the 1k, 2k, 4k, 8k, 16k, 32k, and 64k models, respectively. 
\citet{flamminidotti2019planetary} also investigated the ejection of planetary systems from star clusters. They used nine different models with varying initial virial ratios ($Q=0.4$, 0.5, and 0.6) and different star number densities (500, 5\,000 and 10\,000 stars with $r_{\rm vir} = 1$~pc), comparable to the initial conditions in this study. They studied solar-like planetary systems, excluding Mercury and Venus. The results of planetary system ejections align with this study, as indicated in their table~2. The authors discovered that planetary systems, which are ejected from the star cluster with velocities comparable to the average stellar velocity in the cluster, are less disrupted, compared to those ejected with high velocities \citep[see fig.~12 of][]{flamminidotti2019planetary}. The author argued that multiple soft encounters within the star cluster may increase stellar velocity, and facilitate the ejection of planets. Our work aims to investigate the properties of debris discs born in star clusters, and thus the escaping planetary systems are also included in the statistics below. 

\subsection{The stability of \pdps{}}
\label{sec:stability}

\begin{figure} \centering
    \includegraphics[width=0.47\textwidth]{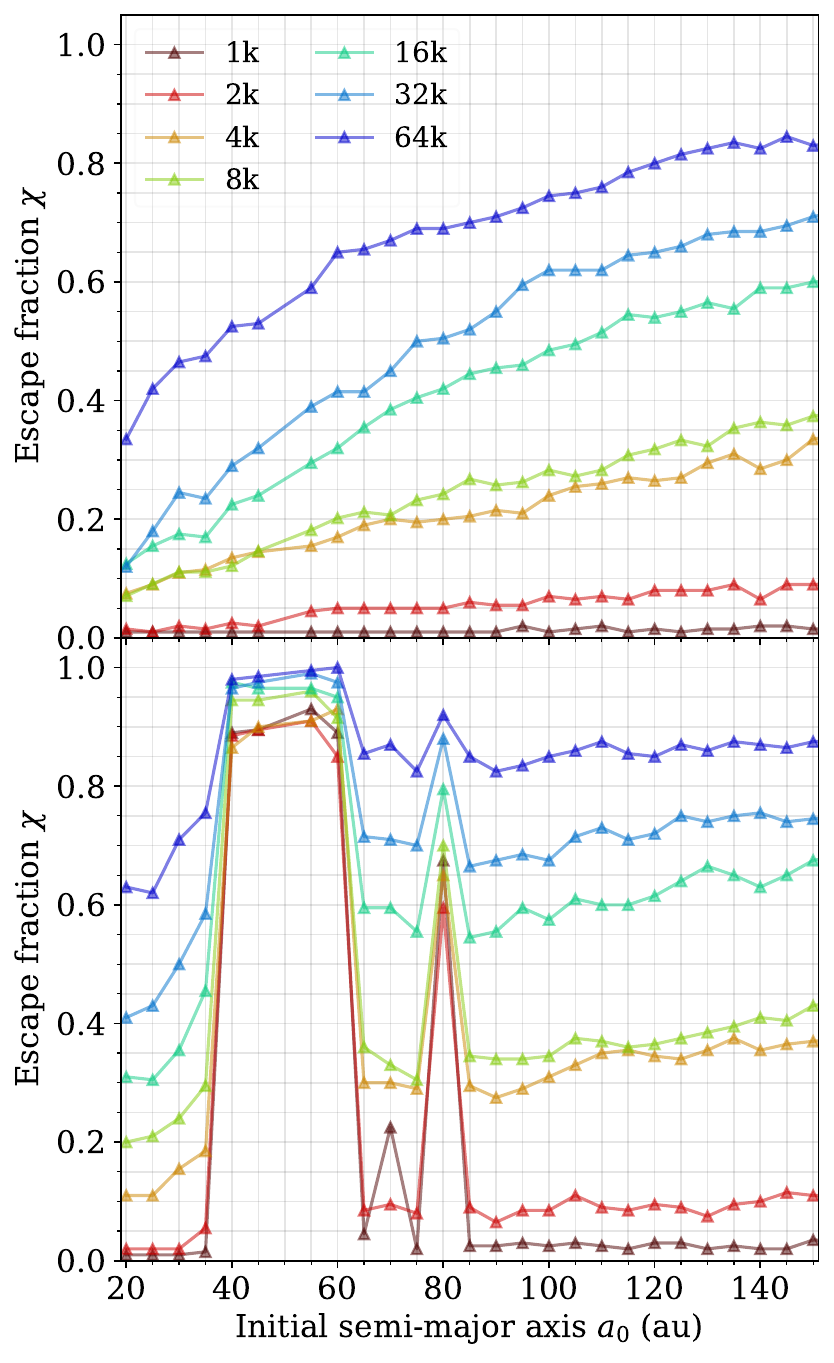} 
   
    \caption[Escape fraction of \pdps{} at 100~Myr as a function of their initial semimajor axis, in different star cluster models, \ensemble{}.]{Escape fraction of \pdps{} at 100~Myr as a function of their initial semimajor axis, in different star cluster models, \ensemble{}. \emph{Top}: P0. \emph{Bottom}: P1. This figure shows results for $a_0 \le 150$~au (the interior region).}
    \label{fig.fesc_over_a0_zoomin}
\end{figure}

\begin{figure} \centering
    \includegraphics[width=0.47\textwidth]{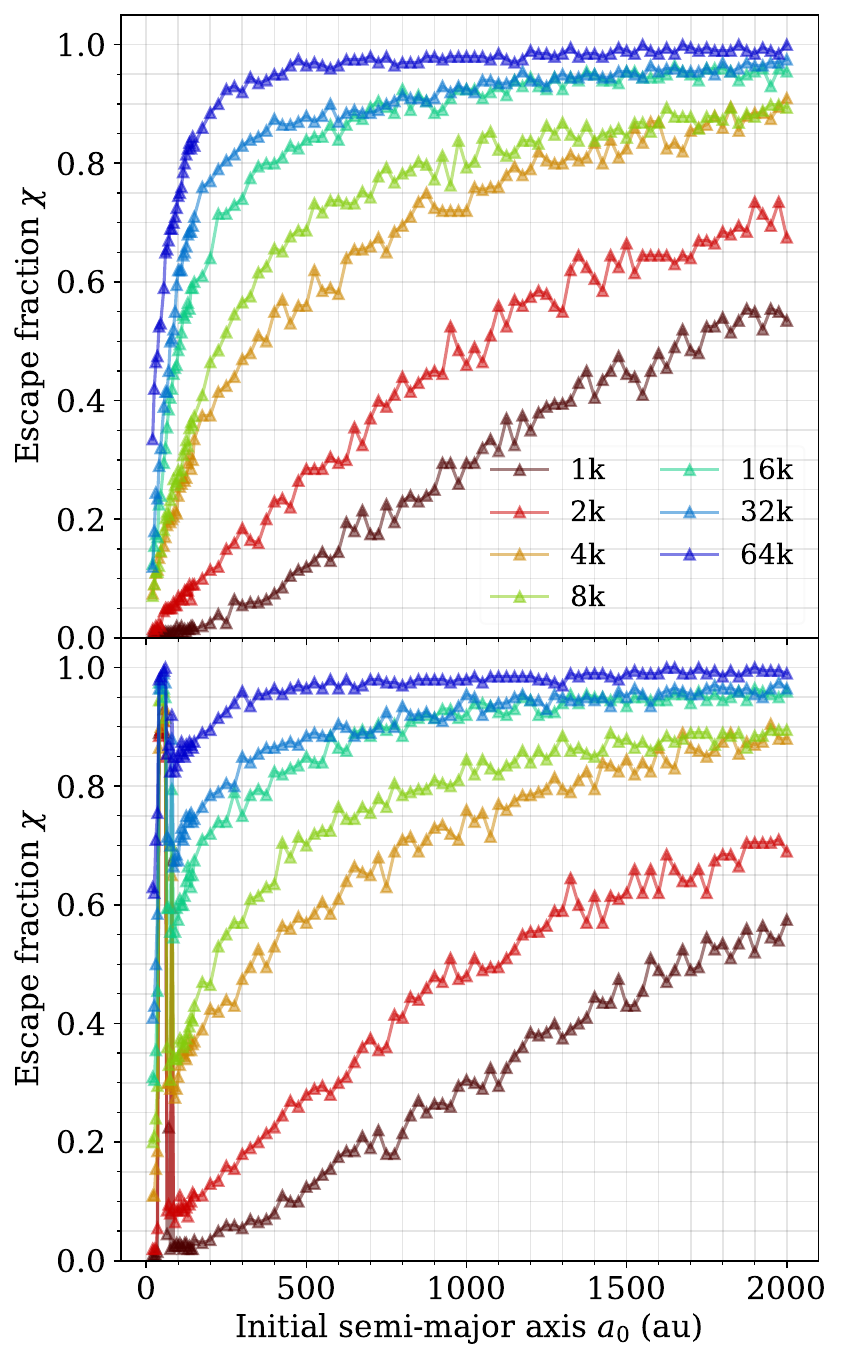} 
   
    \caption{Same as in \fref{fig.fesc_over_a0_zoomin} but for the entire semimajor axis range.}
    \label{fig.fesc_over_a0}
\end{figure}

\begin{figure} \centering
    \includegraphics[width=0.47\textwidth]{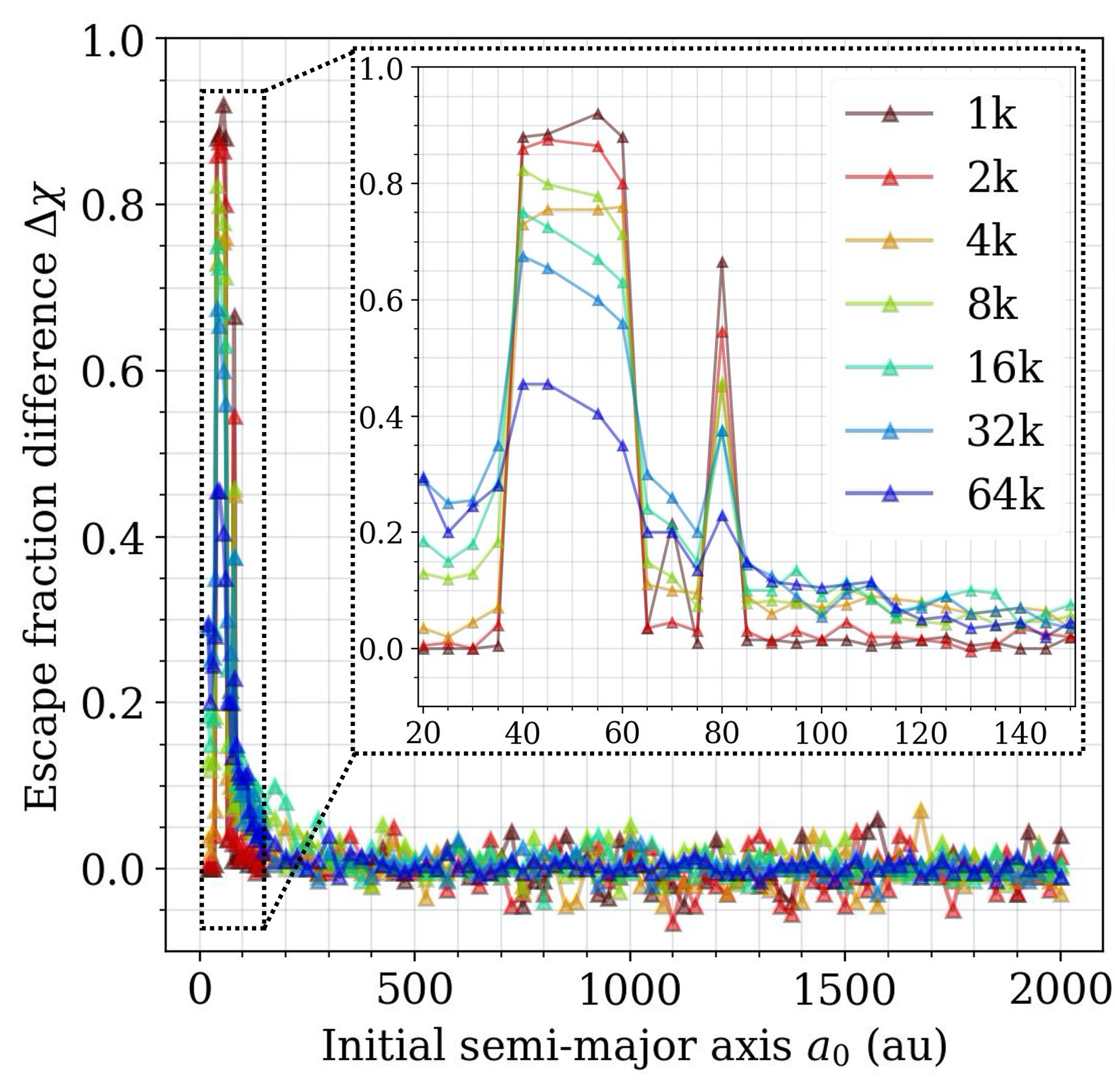} 
   
    \caption{Difference in the survival fraction as a function of initial semimajor axis, in different star cluster models, \ensemble{}, at the end of the simulations ($t=100$~Myr).}
    \label{fig.fesc_over_a0_diff}
\end{figure}

\begin{figure} \centering
    \includegraphics[width=0.47\textwidth]{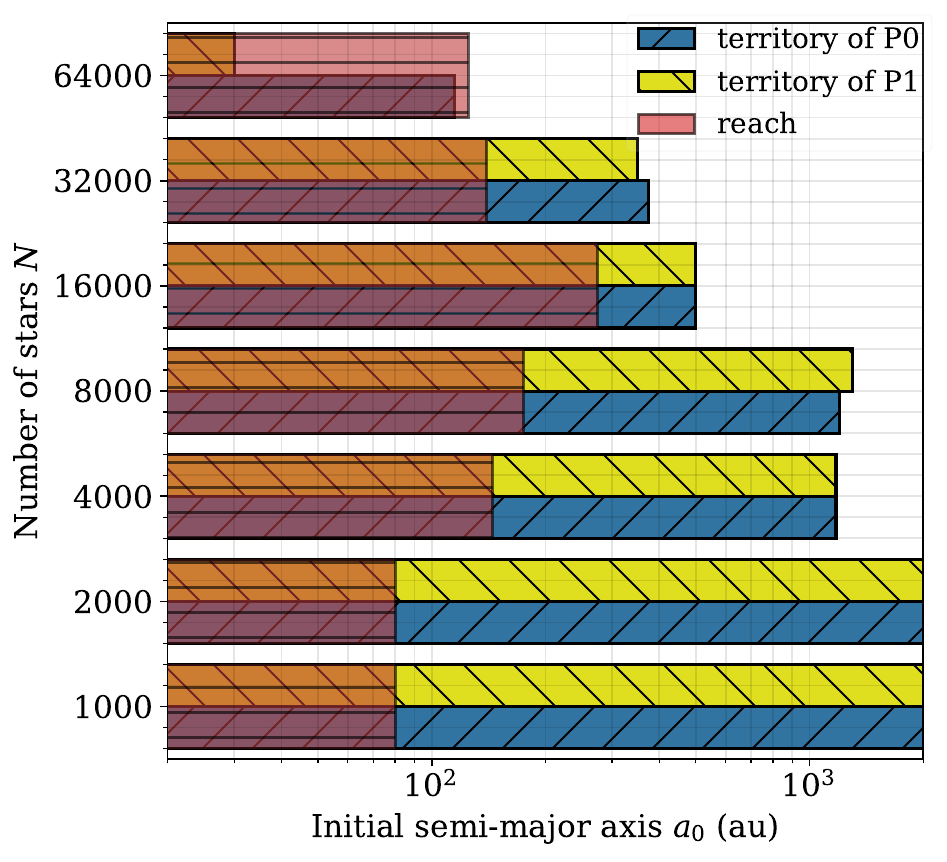}
    \caption{\textit{Reach} and \terri{} of the planetary systems in different star cluster environments.}
    \label{fig.terri_reach_barplot}
\end{figure}

\begin{figure*} \centering
    \includegraphics[width=\textwidth]{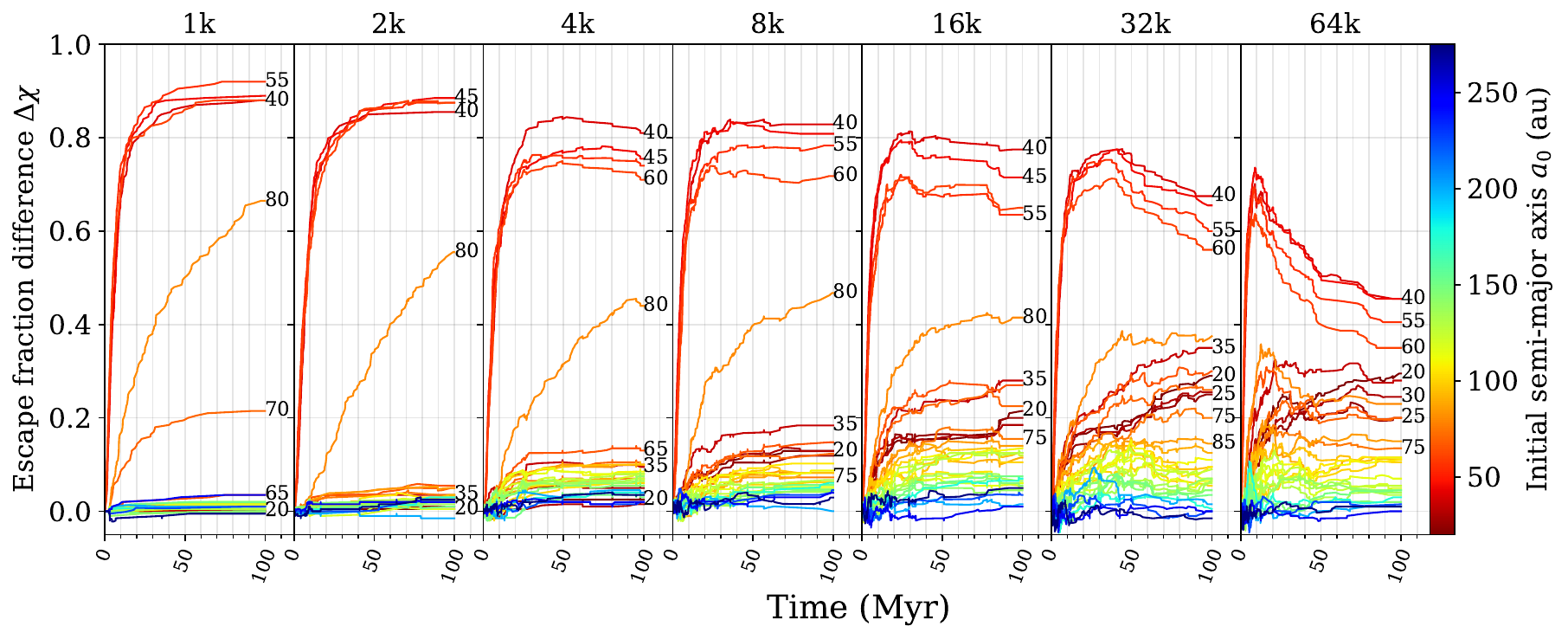} 
    \caption[Difference of survival fraction as a function of simulation time, \ensemble{}.]{Difference of survivor fraction as a function of simulation time, \ensemble{}. Only \pdps{} with $a_0\le275$~au are included, because the maximum \rea{} is 275~au. Different colours indicate different initial semimajor axes. To distinguish curves with similar colours, we annotate values of several of the semimajor axes.}
    \label{fig.diffNesc_over_t_by_a0}
\end{figure*}

The fraction of escaping \pdps{} over time provides a measure for their stability. \fref{fig.fesc_over_a0_zoomin} and \ref{fig.fesc_over_a0} show the escape fraction, $\chi(a_0)$, of \pdps{} as a function of initial semimajor axis, $a_0$. The top panels indicate that denser clusters result in a higher escape fraction for particles of all $a_0$. In the densest (64k) cluster, particles with $a_0>400$~au have escape fractions above 95\% at 100~Myr. 
Escape fractions in all star clusters increase with increasing $a_0$ (\fref{fig.fesc_over_a0}). For clusters with $N \ge 8\,000$, the escape fractions reach 95\% at certain $a_0$. This indicates different planetary system boundaries in different star clusters. By comparing the top and bottom panels of \fref{fig.fesc_over_a0_zoomin} and \ref{fig.fesc_over_a0}, we can see the strong influence of the planet on the interior region ($a_0 \leq 150$~au), while its effect on the exterior region ($a_0 > 150$~au) is small. 

The differences in the escape fraction between the simulation with and without a planet, $\Delta\chi(a_0) \equiv \chi_\mathrm{P1}(a_0) - \chi_\mathrm{P0}(a_0)$, can be used to characterise the influence of the planet. \fref{fig.fesc_over_a0_diff} shows $\Delta\chi(a_0)$ for the different star clusters. The planet's influence on the outer regions of a planetary system is small. In the 8k model of \citet{wu2023influence}, the planet does not influence the region $>400$~au. With our current study we now also demonstrate that the same trend is present for star clusters with other $N$. 

\fref{fig.fesc_over_a0_diff} shows that the planet most influences the region at $40 \leq a_0/\mathrm{au} \leq 60$ and $a_0=$~80~au. The latter region surrounds the 2:1 mean-motion resonance orbit of the planet at $a_0=50$~au. In these two regions, $\Delta\chi$, which indicates the influence of the planet, decreases as $N$ increases. Other than the two regions, $\Delta\chi$ increases mildly with increasing $N$.

For $a_0>400$~au, $\Delta \chi$ oscillates around zero in \fref{fig.fesc_over_a0_diff}. Clusters with larger $N$ show a smoother curve and weaker oscillations, because the escape fraction in this region tends to zero in both the P0 and P1 simulations. In low-mass clusters, on the other hand, the oscillations are significant, due to low-number statistics.

To better describe the characteristics of the planetary system in star clusters, \citet{wu2023influence} defined three regions: the \private{}, the \reach{}, and the \territory{} (hereafter, \pri{}, \rea{}, and \terri{}). In \citet{wu2023influence} the \private{} is $40-60$~au, where the planet clears all particles. In our current study, we find a consistent range of the \private{} across different star cluster densities (lower panel of \fref{fig.fesc_over_a0_zoomin}). We are interested in how the \rea{} and the \terri{} vary in different star cluster densities. For better comparison, we define them quantitatively based on \citet{wu2023influence}:
\begin{equation}
    reach: \mathrm{the~upper~bound~of}~a_0\mathrm{~where}~\Delta\chi \geq 5\%
    \label{eqn.reach_define}
\end{equation}
and
\begin{equation}
    territory: \mathrm{the~upper~bound~of}~a_0\mathrm{~where}~\chi_\mathrm{lim} \geq 90\% \quad .
    \label{eqn.terri_define}
\end{equation}
Here, $\chi_\mathrm{lim}$ is defined as the estimated asymptotic escape fraction \citep{wu2023influence}. Although according to the encounter estimation in \sref{sec.result_star_cluster}, planetary systems in the 16k, 32k, and 64k clusters will still experience many encounters after 100~Myr, but $\chi_\mathrm{lim}$ provides an acceptable estimate of the final stability through fitting and extrapolation. 

\fref{fig.terri_reach_barplot} summarizes the dependence of the \terri{} and \rea{} on the surrounding cluster environment. \terri{} decreases as cluster density increases. The planet has no significant impact on the \terri{}, except in the densest 64k models, where the \terri{} becomes smaller than the \rea{}, and is thus strongly affected. 
The \rea{} remains in the minimum value of around 80~au in small clusters (1k - 2k). This is similar to the isolated model (figure~3 in \citet{wu2023influence}), where the planet only increases the escape fraction for \pdps{} with $a_0 \leq 80$~au. It is thus likely that the effect of the planet and the cluster are independent in sparse clusters. As the cluster density increases, the \rea{} first increases and then decreases, with a maximum value of 275~au for the 16k model. 

\fref{fig.diffNesc_over_t_by_a0} shows $\Delta\chi(t)$ within the \rea{}. In the low-mass star clusters (1k - 8k), the influence of the planet becomes more important with time for all $a_0$. 
In the massive clusters (16k - 64k), the tendency varies with different $a_0$. For particles with $20~\mathrm{au} \leq a_0 \leq 35~\mathrm{au}$, which are inside the planetary orbit, but relatively far from $a=50$~au, the influence of the planet increases with time. For other orbits within the \rea{}, the influence tends to first increase and then decrease with time, especially in 64k model, where the influence decrease with time for most $a_0$.
The planet's influence on particles in the vicinity (40 - 60~au) increases sharply during the first 10~Myr of evolution in all models.
As a result, the most influenced orbits are the planet's vicinity in all models.
In summary, when the stellar density increases, (i) the planet's influence in its vicinity decreases; (ii) the planet's influence in the 2:1 mean-motion resonance orbit decreases; and (iii) the planet's influence in other regions of the planetary system increases.

\subsection{Properties of escaping \pdps{}}

\begin{figure*} \centering
    \includegraphics[width=0.43\textwidth]{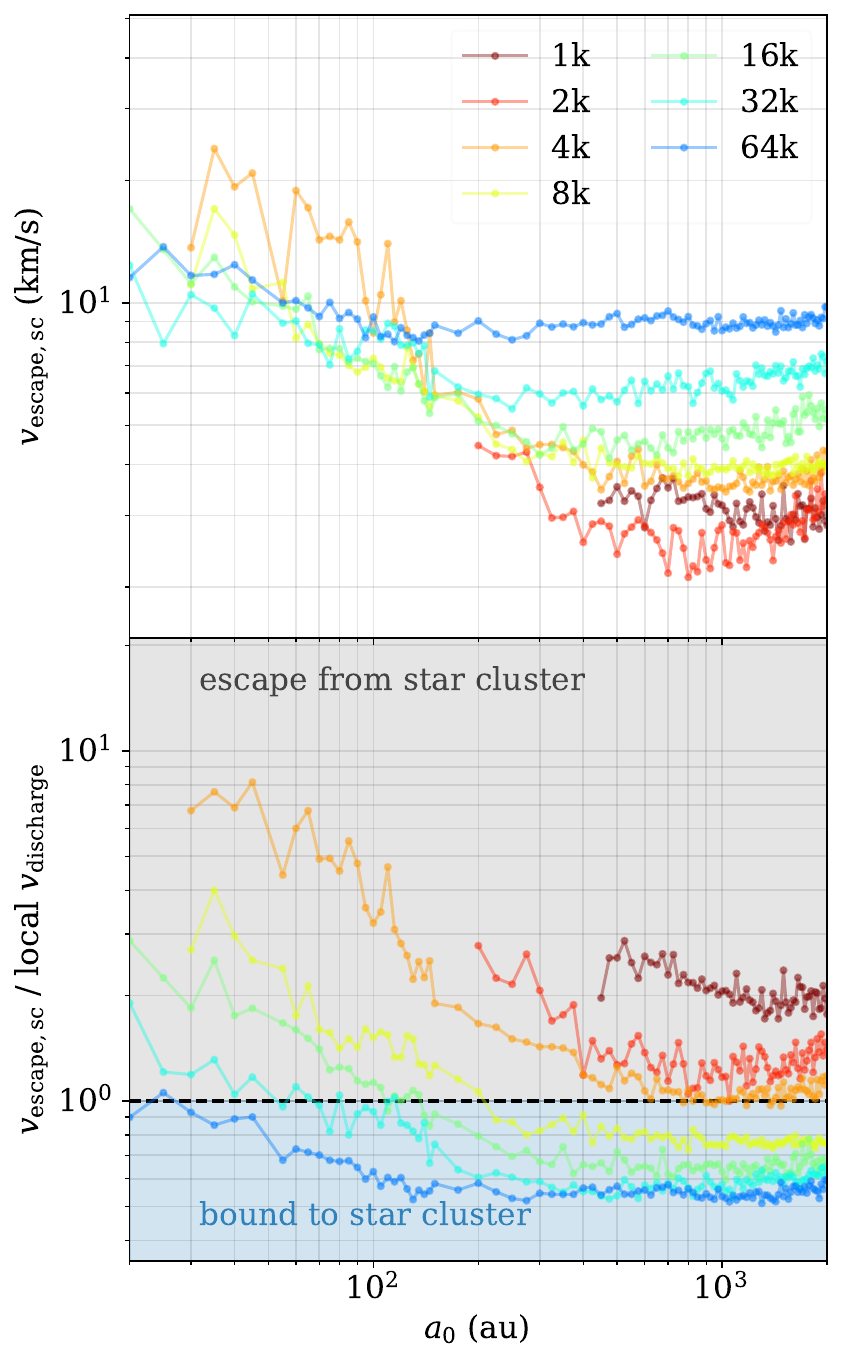} \
    \includegraphics[width=0.43\textwidth]{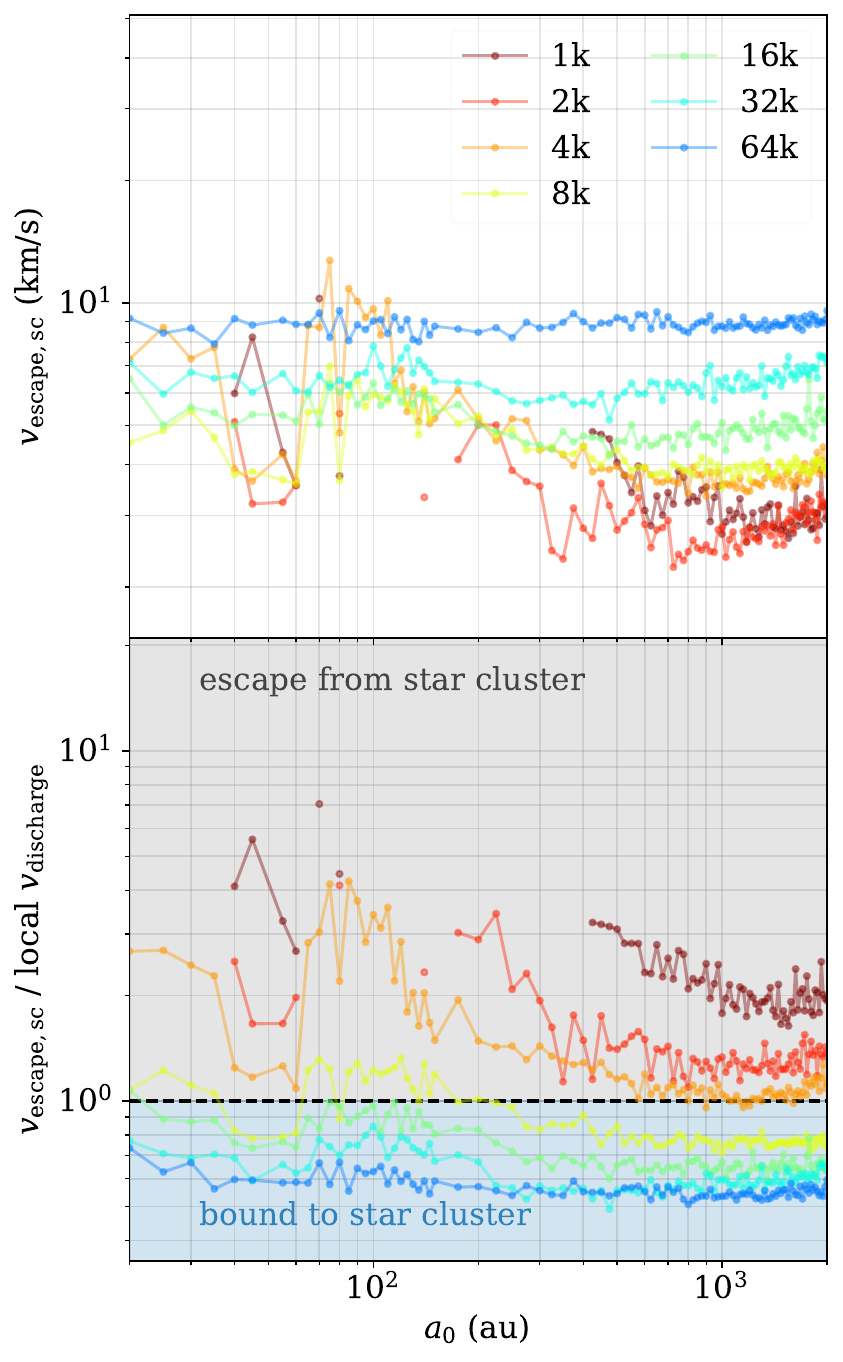} 
    \caption[Statistics of escapers, \ensemble{}, at the end of the simulations.]{Statistics of escapers, \ensemble{}, at the end of the simulations ($t=100$~Myr). Different colours indicate planetary systems in different star cluster models. \emph{Top}: escape speed as a function of initial semimajor axis. \emph{Bottom}: destination factor, i.e., escape speed divided by the local discharge barrier speed of the star cluster. \emph{Left}: P0. \emph{Right}: P1. The solid dots show the median values. Large fluctuations due to low-number statistics are present when the escape fraction is less than 10\%; we therefore omit these data points.  We identify the data representing particles escaping from the star cluster and bound to the star cluster with grey and blue shaded regions, respectively.}
    \label{fig.escaper}
\end{figure*}

A planet or debris particle is considered as having escaped from its planetary system when its eccentricity is larger than 0.99. When such a body escapes has a low speed, it becomes a free-floating planet or debris particle in the star cluster. If it has a sufficiently high speed, it may immediately leave the star cluster.

Characterizing the kinematics of the escaping debris particles in near star clusters facilitates efforts aimed at finding the origin of objects such as 1I/‘Oumuamua \citep{oumuamua} and CNEOS 2014-01-08 \citep{CNEOS2014}. Since the particles in the P0 simulations do not interact with each other, each of their instances can also be considered as a single planet in an exoplanetary system. Therefore, such results can be applied to some extent to the physical properties of free-floating planets in clusters as well. The first paper of this series, \citet{wu2023influence}, explored the relationship between the speeds of escaping particles and their initial semi-major axes in planetary systems. Below, we further investigate the effect of different cluster environments on the kinematics of escaping particles.

Following the definitions in equations (9) - (13) of \citet{wu2023influence}, we calculate the escape speed ($v_\mathrm{escape,sc}$) and the local discharge barrier speed of the star cluster (local $v_\mathrm{discharge}$). In \fref{fig.escaper}, we show the statistics of the escape speed and the destination factor (i.e., the ratio between $v_\mathrm{escape,sc}$ and the local $v_\mathrm{discharge}$), as a function of initial semimajor axis. 

In the P0 simulation, the escape speed decreases with increasing $a_0$ in all star cluster models. As a result, \pdps{} with smaller $a_0$ are more likely to be discharged from the star cluster and thus become interstellar objects. \Pdps{} with larger $a_0$ are more likely to remain bound to their host star cluster. 

The bottom left panel of \fref{fig.escaper} clearly shows that the destination factor decreases for clusters with higher stellar densities. Dense clusters are likely to retain most particles that escape from planetary systems. In sparse clusters, on the other hand, such particles often leave the star cluster immediately. The sparsest (1k and 2k) clusters hardly retain any escapers and are thus essentially devoid of free-floating planets and debris. Note that only median values are presented in \fref{fig.escaper}; variation is present. For example, for the 2k cluster, approximately half of the particles with $a_0 \simeq 1000$~au remain in the cluster. 
The central densities of our 1k and 2k clusters are comparable to those of many observed open clusters, such as the Pleiades (see \sref{paper2.method.sec}). These sparse open clusters, which resemble our simulated models, are thus unlikely to retain (1) escaping planetary debris particles from planetary systems without planets and (2) escaping exoplanets from planetary systems that initially have only one planet. 
\change{In denser clusters, particles tend to escape from their planetary systems with higher velocities, but are also more likely to remain members of the star cluster. This is presumably due to the deeper potential wells of denser clusters that prevent particle discharge.}

When we introduce a planet into the simulations, the escape speeds generally decrease. The influence of the planet's presence strongly depends on the initial orbit ($a_0$) of an escaper. 
The influence of the planet's presence is most prominent in the \pri{} $40-60$~au. In this region, the escape speeds decrease significantly. For these escapers in the 64k model, the escape speed is 10\% higher in the P0 simulations than in the P1 simulations, while in the 4k model it is a factor $\sim50$ higher. The sparser the star cluster, the greater the decrease in escape speed for escapers in the \pri{} region.

A comparison between left-hand and right-hand panels of \fref{fig.escaper} indicates that the planet has no significant influence on $a_0 > 200$~au. Escapers affected by the planet, but not present in the \pri{} region, are initially located in $20 \leq a_0/\mathrm{au} < 40$ and $60 < a_0/\mathrm{au} \leq 200$. For these escapers, $v_\mathrm{escape,sc}$ are lower for higher cluster densities. We also find that the planet has a similar degree of influence on debris particles in all clusters: the ratio of $v_\mathrm{escape,sc}$ in P1 to that in P0 at certain $a_0$ is very similar for all clusters. 

In the P1 model, the destination factor is lower than that of the P0 model, as a result of the decrease in the escape speed. With planets present in the planetary systems, more escapers tend to remain in the star cluster, and the destination factor decreases as the cluster density goes up.

\subsection{The architecture of remaining debris discs}
\label{sec:survive}

\begin{figure*} \centering
    \includegraphics[width=\textwidth]{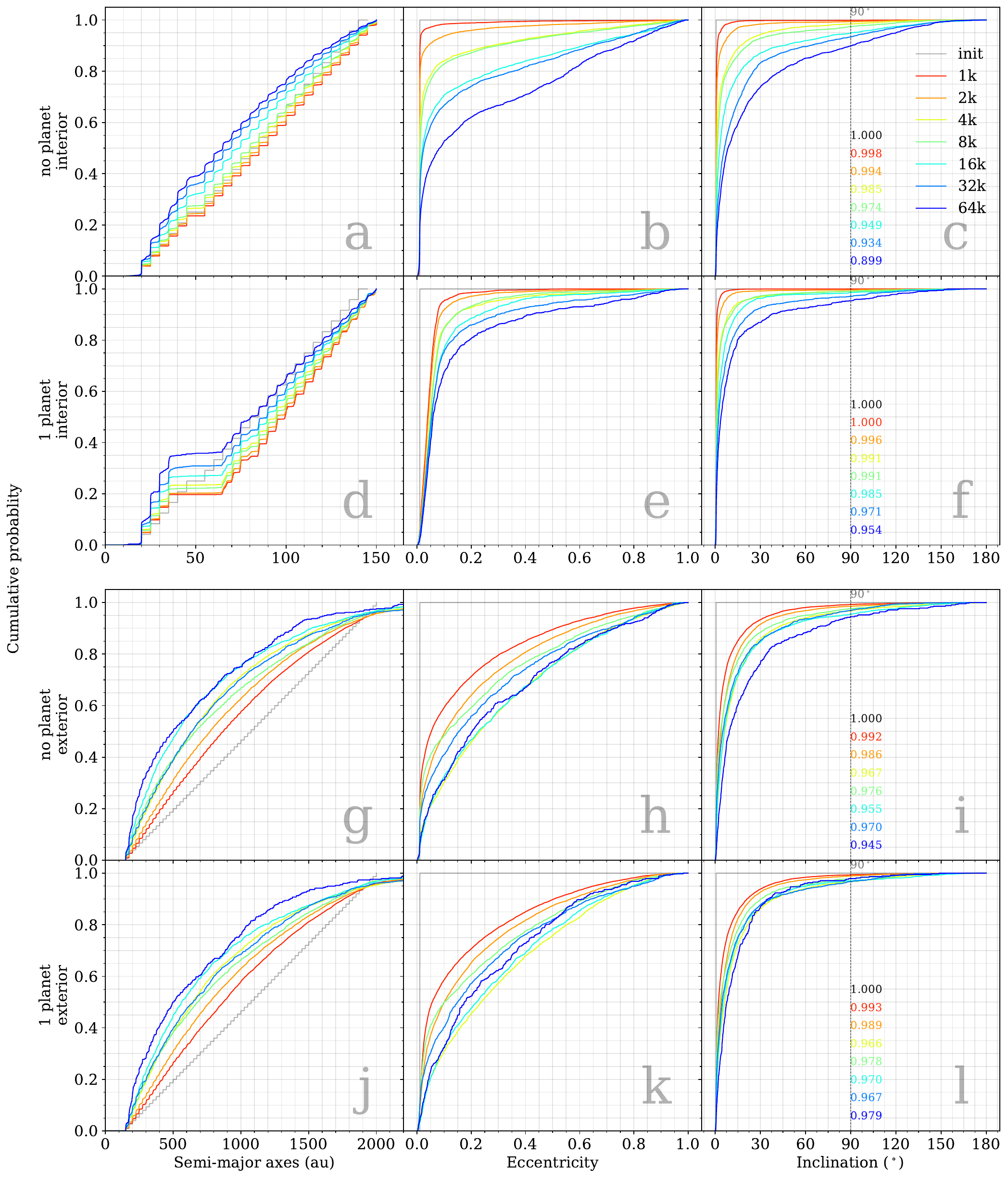} 
    \caption[Cumulative distribution functions of semimajor axes, eccentricity and inclination of surviving \pdps{} at $t = 100$~Myr in different star cluster models, \ensemble{}.]{Cumulative distribution functions of semimajor axes (left-hand panels), eccentricity (middle panels) and inclination (right-hand panels) of surviving \pdps{} at $t = 100$~Myr in different star cluster models, \ensemble{}. Different colours represent different star cluster models, while the grey curve indicates the initial conditions of the planetary systems. Results are shown for the region $a_0 \le 150$~au (top panels) and $a_0>150$~au (bottom panels), respectively. In the right-hand panels, the probability of particles with prograde orbits are annotated for different star clusters.}
    \label{fig.survivor}
\end{figure*}

\begin{figure*} \centering
    \includegraphics[width=\textwidth]{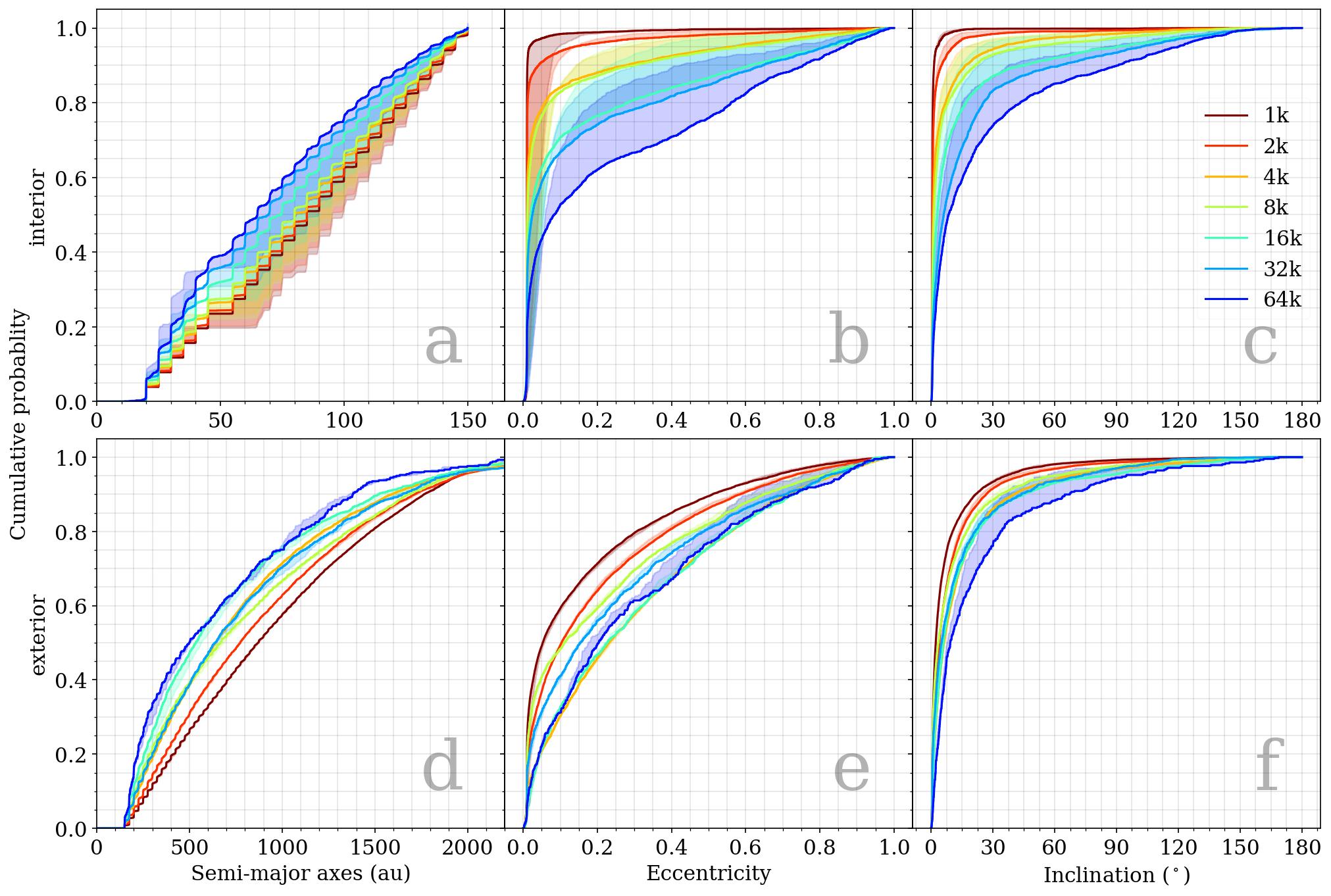} 
    \caption[Same data as \fref{fig.survivor}, but P0 and P1 data are overlapped for better comparison.]{Same data as \fref{fig.survivor}, but P0 and P1 data are overlapped for better comparison. The shaded area indicates the difference between the P1 simulations (thick-coloured curves) and the P0 simulations.}
    \label{fig.survivor_1on0}
\end{figure*}

\begin{figure*} \centering
    \includegraphics[width=0.712\textwidth]{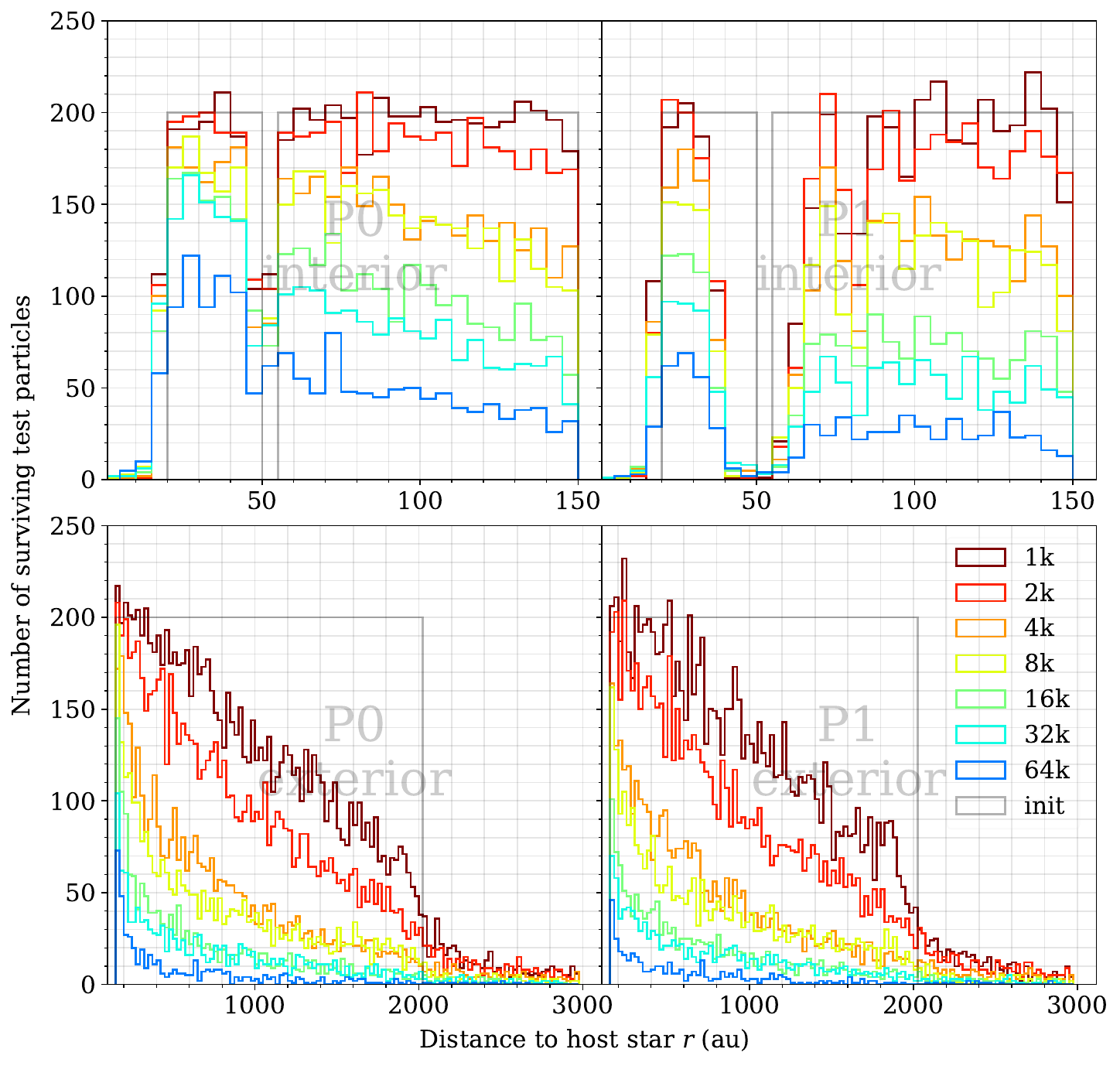}
    \caption[3D radial distribution of surviving \pdps{}, \ensemble{} at $t = 100$~Myr.]{3D radial distribution of surviving \pdps{}, \ensemble{} at $t = 100$~Myr. \emph{Top}: interior region ($r \le 150$~au). \emph{Bottom}: exterior region ($r > 150$~au). \emph{Left}: P0. \emph{Right}: P1.
    }
    \label{fig.shape}
\end{figure*}

\begin{figure*} \centering
    \includegraphics[width=\textwidth]{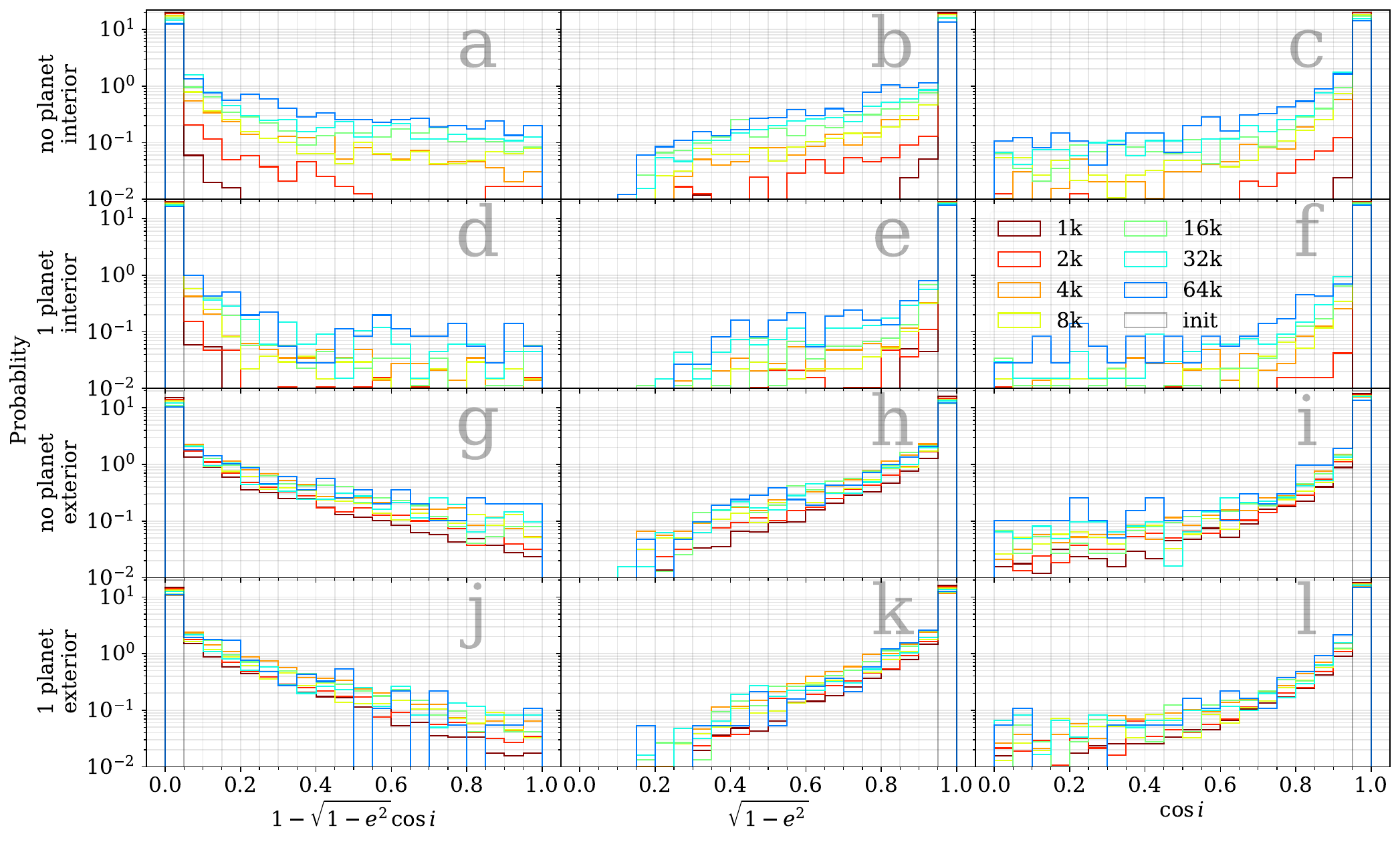} 
    \caption[Normalized histogram of \namdk{}, $\sqrt{1-e^2}$, and $\cos i$ for surviving \pdps{} at $t = 100$~Myr.]{Normalized histogram of \namdk{}, $\sqrt{1-e^2}$, and $\cos i$, for surviving \pdps{} at $t = 100$~Myr, \ensemble{}.}
    \label{fig.survivor_angular_momentum}
\end{figure*}

Below, we will discuss the properties of the remaining \pdps{}: the orbital element distributions, the influence of the star cluster environment, the impact of the planet, the structure of the debris disc, and the angular momentum distribution of the \pdps{}.

\subsubsection{Orbital element distributions}

The distribution of the semi-major axis, eccentricity, and inclination at $t=100$~Myr of surviving particles is shown in \fref{fig.survivor}. For the interior region, we see that \pdps{} in denser star cluster models have smaller semimajor axes (\fref{fig.survivor}a, d). As the cluster density increases, the eccentricity and inclination distributions in the interior region become more uniform (\fref{fig.survivor}b, c, e, f). In the P0 simulations, some surviving particles have near-circular orbits (\fref{fig.survivor}b). In the P1 simulations, the distribution of survivors is more or less uniform between $e=0$ and $e\sim0.1$ (\fref{fig.survivor}e). For example, about 90\% survivors in the 2k, P1 models have a uniform distribution in the eccentricity range $e=0-0.08$. This may be partly caused by the planet's non-zero  eccentricity.

In the exterior region, there does not appear to be a simple relation between the orbital elements ($a$, $e$ or $i$) and cluster density. There is also no clear correlation between the evolution of the orbits and the presence of the planet. When not considering the 4k and 16k models from the 3rd and 4th row of \fref{fig.survivor}, the semi-major axes of surviving \pdps{} decrease, eccentricity and inclination increase as cluster density increases. Further investigation is needed to determine whether or why the orbital element distributions of the surviving particles in the 4k and 16k models behave differently.

By comparing the properties of the surviving \pdps{} in the exterior and interior regions, we see in the exterior region the eccentricity is closer to a uniform distribution (\fref{fig.survivor} middle column) and the inclination is also generally larger (\fref{fig.survivor} right-hand column). 

In \fref{fig.survivor} right-hand column, we identify the fraction of surviving \pdps{} with inclinations larger than 90$^\circ$, i.e., particles that obtain retrograde orbits. When the planet is present, the inclination of a surviving particle is generally lower than that of the P0 simulation.

The influence of the planet is shown in \fref{fig.survivor_1on0}. The bold curve indicates the P0 simulation, and the other margin of the shaded area indicates the P1 simulation. The difference between P0 and P1 simulations is shown as the shaded area. 

In \fref{fig.survivor_1on0}b, except in the 64k model, we see the presence of the planet reduces the number of two groups of particles: (1) eccentricity too low (close to zero), as the shaded area is on the right-hand side of the bold curve in the low-eccentricity region and (2) eccentricity too high, since the shared area is on the top of the bold curve in the region of $e>0.2$. 

\fref{fig.survivor_1on0}c shows that the overall inclination is lower in P1 simulations than that in P0 in the interior region. It may be possible that the planet is removing high-inclination particles. Since we initially set the planet on the reference plane and thus has zero inclination, this is a possible consequence of the planet exchanging angular momentum with \pdps{} whose inclination is raised by stellar flybys and stabilizing them. The mechanism behind this can be addressed in future studies. 

The planet has no significant impact on the exterior region. The most notable difference is on \pdps{} in the 64k model, where $e$ and $i$ are overall lower than P1, which is consistent with the phenomenon in the interior region.

\citet{wu2023influence} argued that the planet may accelerate the evolution of planetary systems in clusters, because the CDFs of the eccentricity and inclination at 10~Myr and 100~Myr are closer to each other in the P1 simulation. We examine the survivor's CDF at different simulation times \citep[like figure~16 of][]{wu2023influence} in each cluster model. The same phenomenon is also present in all our star cluster models except 64k. 

\subsubsection{Radial disc profile} 

\fref{fig.shape} shows the distribution of distances between the \pdps{} and the host star at $t=100$~Myr, indicating the structure of the remaining disc. Note that the final distribution depends strongly on the choice of the initial conditions. \fref{fig.shape} shows the combined result for all planetary systems, although strong variations are present for individual systems. The ensemble's radial profile in the exterior region shows a nonlinear decrease with $r$. 
\change{We explore two functions that can be used to accurately describe the radial profile: the commonly adopted two-parameter power-law distribution, $\rho = \rho_0 r^{\alpha}$ \citep[e.g.,][]{wyatt2008ddrev,armitage2011dynamicprotorev} and (ii) the two-parameter exponential decay function $\rho(r) = \rho_0 e^{r/r_0}$. Here, $\rho_0$, $\alpha$, and $r_0$ are fitting parameters. We fit both functions to the distribution and obtain the optimal fitting parameters\footnote{Using \texttt{scipy.optimize.curve\_fit} in \texttt{Python}}. Based on both the $\chi^2$ and BIC (Bayesian Information Criterion) tests, the exponential decay function provides a better description of the radial disc profile in the 1k$-$4k models, while the power-law function suits the profile in other models.}

\subsubsection{Angular momentum deficit of surviving particles} 

The evolution of the planetary systems can be further investigated using the concept of angular momentum deficit \citep[AMD;][]{laskar1997large,laskar2000on-the-spacing,laskar2017amd-stability}. The circular angular momentum (CAM) represents the maximum possible angular momentum of a particle $k$ in orbit around a star, 
\begin{equation}
    \mathrm{CAM}_k=m_k \sqrt{a_k G M_{\star}} \quad ,
\end{equation}
where $m_k$ is the particle mass, $a_k$ is the semimajor axis, $G$ is the gravitational constant, and $M_{\star}$ is the host star mass. The angular momentum deficit of the particle is
\begin{equation}
    \mathrm{AMD}_k=m_k \sqrt{a_k G M_{\star}}\left(1-\sqrt{1-e_k^2} \cos i_k\right) \quad ,
\end{equation}
where $e_k$ and $i_k$ are the eccentricity and inclination, respectively. Since the \pdps{} in our simulations are massless, we use the normalized AMD \citep{chambers2001making,turrini2020normalized}:
\begin{equation}
    \mathrm{NAMD}_k = \frac{\mathrm{AMD}_k}{\mathrm{CAM}_k} = 1-\sqrt{1-e_k^2} \cos i_k \quad .
\end{equation}
Since all our particles are initially nearly coplanar and on nearly circular orbits, their initial \namdk{} is close to zero. 

\fref{fig.survivor_angular_momentum} shows the distribution of \namdk{}, the contribution from the eccentricity, $\sqrt{1-e^2}$, and the contribution from the inclination, $\cos i$. Although the planet affects the eccentricity of exterior survivors, the planet hardly affects  their \namdk{}. In the P1 case, as a result of lower $e$ and $i$ compared with P0, $\sqrt{1-e^2}$ and $\cos i$ are larger, resulting in a smaller \namdk{}. \citet{turrini2020normalized} argue that the distribution of \namdk{} can show signs of the dynamical history of the planetary systems. The \namdk{} would thus suggest that surviving \pdps{} P1 simulations are less perturbed, but we do not observe this. The P1 simulation has an overall higher escaper fraction than P0 (\sref{sec:stability}), which means the \pdps{} are more strongly perturbed.

\section{Discussion and Conclusions}
\label{paper2.conclusions.sec}

Stars and planetary systems are believed to originate in clustered star-forming regions. The early formation of planetary systems is thus affected by their stellar environment. Close encounters with neighbouring stars can leave dynamical imprints on the planetary architecture of such systems and their debris discs. In this study, we investigate how debris discs dynamically evolve under the influence of neighbouring stellar populations and the presence or absence of a 50~au planet. We use the \lpsp{} code for combining the codes \nbo{} and \reb{} to evolve debris disc structures that are initially located between $20-2000$~au to the host stars in star clusters containing 1\,000 to 64\,000 stars, and compare between the simulations with and without a Jupiter-mass planet initially at 50~au. Our main findings can be summarized as follows:

\begin{enumerate} 

    \item We have derived a general expression of the encounter rate of planetary systems in star clusters \erefp{eqn.encounter_rate}. \change{The number of stellar encounters estimated using the expression is in good agreement with the direct measurements from the simulations.} This expression provides a useful tool for estimating the number of encounters experienced by a typical star in a star cluster.

    \item More \pdps{} are ejected from the planetary systems in denser star clusters. To quantify the star cluster's influence on the \pdps{}, we define the \territory{} (equation~\ref{eqn.terri_define}; \fref{fig.terri_reach_barplot}). The \terri{} decreases as cluster density goes up. The presence of a planet in the system has no significant effect on \terri{}, except in the densest 64k model.

    \item To quantify the planet's influence on the \pdps{}, we define \reach{} (equation~\ref{eqn.reach_define}; \fref{fig.terri_reach_barplot}). With the increasing cluster densities, the \rea{} first increases and then decreases. The maximum \rea{} occurs at 275~au in the 16k model.

    \item As the cluster density increases, the influence of the planet on the \pdps{} in its vicinity, as well as those in the 2:1 mean-motion resonance orbit, decreases. The influence of the planet on the \pdps{} in the other regions, however, increases. 
    
    \item For particles escaping from the planetary system, escaping particles that are initially near the host star are more likely to be ejected from the star cluster in all our cluster models. Escapers with larger $a_0$ are more likely to remain part of the cluster (\fref{fig.escaper}). The cluster's ability to retain planetary escapers increases as cluster density increases. The presence of a planet tends to reduce the escape speed of escapers, and thus increases the probability of these \pdps{} to remain gravitationally bound to the cluster.
    
    \item As the star cluster density increases, the eccentricity and inclination distribution of surviving \pdps{} tends to increase. A comparison between the P1 and P0 simulations indicates that the $e$ or $i$ of \pdps{} are generally lower when the planet is present.

    \item The radial distribution of our remaining debris discs \change{can be well described with exponential functions in sparse clusters (1k$-$4k models), and with power-laws in dense clusters (8k$-$64k models).}
    
\end{enumerate}

A deeper understanding of the early formation and evolution of planetary systems would require modelling gas in both planetary systems and star clusters \citep{lada2003embedded,perryman2018exoplanet-handbook}, and modelling the photo-evaporation process of the disc. The latter may greatly reduce the disc mass, especially in the hot region of a protoplanetary disc \citep{armitage2011dynamicprotorev}. Although our study focuses on solar-mass host stars, planetary systems around more massive stars may be more vulnerable to dynamical disruption \citep{hands2019the-fate,stock2020resonant}, primarily because they tend to sink to the star cluster centre due to mass segregation. In the outskirts of the star cluster, planetary systems experience fewer close encounters and are more likely to survive as a whole \citep[e.g.,][]{cai2019on-the-survivability}. Similar to \citet{nesvold2015a-smack}, we study models with a Jupiter-mass planet at $a_0=50$~au, as giant planets usually form in cold regions of planetary systems, and are more easily perturbed by neighbouring stars than habitable-zone planets. In this study, we see that a wide-orbit planet also affects regions closer to the star (\fref{fig.fesc_over_a0_zoomin}). Planetary systems with debris discs in the Galactic field may thus have imprints of their birth environment, if a wide-orbit planet is (or was) present in the system. Our study provides a first look into the dynamical interplay between star clusters, planets, and debris structures. A more comprehensive study is required to obtain more general results. The impact of varying planetary mass, eccentricity, and inclination on debris discs in isolated planetary systems was studied in \citet{nesvold2015a-smack,tabeshian2016detection,tabeshian2017detection}. The joint influence of different planetary configurations and the star cluster environment can be further studied in the future. 

The presence of the planet has the most prominent impact in an intermediate-mass cluster. Our methodology does not allow us to determine the upper limit for the stellar density of a star cluster that can harbour planetary systems. Initially, there are 400 \pdps{} inside $r<30$~au, and even for the planetary systems in the densest 64k cluster model, there are still 289 \pdps{} inside $r<30$~au (\fref{fig.shape}). Our quantified measurement \terri{} (\fref{fig.terri_reach_barplot}) shows that planetary systems in 64k cluster may still harbour particles inside $30$~au (P0) or $115$~au (P1), which is comparable to the location of one of our Solar System's debris discs, the Kuiper belt. It is thus hard to explain the rarity of observed debris discs singularly by the dynamical interaction of the disc and the stellar environment. Nevertheless, the combined effect of the flyby stars and planets is promising to explain the scarcity of observed discs in clusters, and also systems in which planet and disc co-exist. In the P1 simulations of the 64k cluster, only a few \pdps{} survive, and many of these survivors have highly eccentric or inclined orbits. In the 16k model, a single planet at $a=50$~au can affect regions as wide as 275~au and eject many more particles. Many observed planetary systems host more than one planet. The interplanetary interactions along with stellar encounters may be potentially more destructive, removing most debris disc particles and making the remaining structure too sparse to be observed. 

All star cluster models in our study have an initial half-mass radius of 0.78~pc. To obtain more general results, additional $N$-body simulations are required.
Our models did not include primordial binaries or triple systems. In all our star cluster models, the fraction of dynamically-formed binary systems remains below 1\% throughout the 100~Myr simulation. Binaries as perturbing stars have higher masses and can strongly destabilize the planetary systems \citep[e.g.,][]{li-adams2015cross-sections,wangyh2020planetary}. Binaries hosting planetary companions are dynamically complex, and can have different configurations (S-type and P-type). Modelling them is interesting but beyond the scope of the current paper. \citet{umbreit2011disks} found that discs in triple systems have similar surface density distributions as those that experienced two-body stellar encounters, but are generally less massive than the latter.  

The current study provides a benchmark for future studies that will explore the parameter space of star clusters and planetary systems in greater detail.

\section*{Acknowledgements}

We are grateful to the referee, Alexander Mustill, for providing valuable comments and suggestions that helped to improve this paper.
This research was supported by the Postgraduate Research Scholarship (grant PGRS1906010) of Xi'an Jiaotong-Liverpool University (XJTLU). 
KW acknowledges Mingze Sun, Fabo Feng, Hui-gen Liu, Tron Du, and Leonard Benkendorff for their helpful discussions, and acknowledges Yifan Wang, Xiaoying Pang, Xi Chen, and Shujun He for providing computational resources. 
MBNK acknowledges support from the National Natural Science Foundation of China (grant 11573004). 
FFD and RS acknowledge support from the DFG priority program SPP 1992 “Exploring the Diversity of Extrasolar Planets” under project Sp 345/22-1. 
This paper utilizes data from the SVO Stars with debris discs and planets Data Access Service at CAB (CSIC-INTA)\footnote{\url{http://svocats.cab.inta-csic.es/debris2/index.php}}, catalogue of resolved debris discs maintained by Nicole Pawellek and Alexander Krivov\footnote{\url{https://www.astro.uni-jena.de/index.php/theory/catalog-of-resolved-debris-disks.html}} and data from circumstellardiscs.org\footnote{\url{https://www.circumstellardiscs.org/}}. 
This research has made use of the SIMBAD database, operated at CDS, Strasbourg, France \citep{simbad}. 
This paper makes use of the \python{} packages SciPy \citep{scipy}, NumPy \citep{numpy}, Matplotlib \citep{matplotlib}, and Seaborn \citep{seaborn}.

\section*{Data Availability}
The data underlying this article will be shared on reasonable request to the corresponding author.

\bibliographystyle{mnras}
\bibliography{planet_on_comets_hard_link}



\bsp	
\label{lastpage}
\end{document}